\documentclass[fleqn,usenatbib]{mnras}

\usepackage[T1]{fontenc}
\usepackage[utf8]{inputenc}

\DeclareRobustCommand{\VAN}[3]{#2}
\let\VANthebibliography\thebibliography
\def\thebibliography{\DeclareRobustCommand{\VAN}[3]{##3}\VANthebibliography}

\usepackage{graphicx} 
\usepackage{acro}
\usepackage{units}
\usepackage{booktabs}
\usepackage{dcolumn}
\usepackage{makecell}
\usepackage{graphicx}
\usepackage{tabularx}
\usepackage{amsmath,amsfonts,amssymb}
\usepackage{xspace}
\usepackage{mathrsfs}
\usepackage[normalem]{ulem}
\usepackage{multirow}
\usepackage{color}
\usepackage{xcolor}
\usepackage{natbib}
\usepackage{verbatim}
\usepackage{physics}
\usepackage{braket}
\usepackage{colortbl}
\usepackage{newtxtext,newtxmath}
\usepackage[version=3]{mhchem}

\DeclareOldFontCommand{\rm}{\normalfont\rmfamily}{\mathrm}
\DeclareOldFontCommand{\sf}{\normalfont\sffamily}{\mathsf}
\DeclareOldFontCommand{\tt}{\normalfont\ttfamily}{\mathtt}
\DeclareOldFontCommand{\bf}{\normalfont\bfseries}{\mathbf}
\DeclareOldFontCommand{\it}{\normalfont\itshape}{\mathit}
\DeclareOldFontCommand{\sl}{\normalfont\slshape}{\@nomath\sl}
\DeclareOldFontCommand{\sc}{\normalfont\scshape}{\@nomath\sc}
\DeclareRobustCommand*\cal{\@fontswitch\relax\mathcal}
\DeclareRobustCommand*\mit{\@fontswitch\relax\mathnormal}

\DeclareAcronym{EOS}{short=EOS,long=equation of state}
\DeclareAcronym{GW}{short=GW,long=gravitational wave}
\DeclareAcronym{BH}{short=BH,long=black hole}
\DeclareAcronym{NS}{short=NS,long=neutron star}
\DeclareAcronym{PNS}{short=PNS,long=proto-neutron star}
\DeclareAcronym{BHNS}{short=BHNS,long=black hole - neutron star}
\DeclareAcronym{BNS}{short=BNS,long=binary neutron star}
\DeclareAcronym{NSE}{short=NSE,long=nuclear statistical equilibrium}
\DeclareAcronym{LTE}{short=LTE,long=local thermodynamic equilibrium}
\DeclareAcronym{NLTE}{short=NLTE,long=non-local thermodynamic equilibrium}
\DeclareAcronym{KN}{short=KN,long=Kilonova,short-plural-form=KNe,long-plural-form=Kilonovae}
\DeclareAcronym{SN}{short=SN,long=Supernova,short-plural-form=SNe,long-plural-form=Supernovae}
\DeclareAcronym{CCSN}{short=CCSN,long=core-collapse Supernova,short-plural-form=CCSNe,long-plural-form=core-collapse Supernovae}
\DeclareAcronym{MRSN}{short=MRSN,long=magnetorotational Supernova,short-plural-form=MRSNe,long-plural-form=magnetorotational Supernovae}
\DeclareAcronym{SE}{short=SE,long=stripped-envelope}
\DeclareAcronym{EM}{short=EM,long=electromagnetic}
\DeclareAcronym{IR}{short=IR,long=infrared}
\DeclareAcronym{GRB}{short=GRB,long=gamma-ray burst}
\DeclareAcronym{ZAMS}{short=ZAMS,long=zero-age main sequence}
\DeclareAcronym{WR}{short=WR,long=Wolf-Rayet}
\DeclareAcronym{BL}{short=BL,long=broad-line}
\DeclareAcronym{SNR}{short=SNR,long=signal-to-noise ratio}
\DeclareAcronym{VLT}{short=VLT,long=Very Large Telescope}
\DeclareAcronym{ESO}{short=ESO,long=European Sourthern Observatory}
\DeclareAcronym{JWST}{short=JWST,long=James Webb Space Telescope}
\DeclareAcronym{E1}{short=E1,long=electric dipole}
\DeclareAcronym{M1}{short=M1,long=magnetic dipole}
\DeclareAcronym{E2}{short=E2,long=electric quadrupole}
\DeclareAcronym{NIST}{short=NIST,long=National Institute of Standards and Technology}

\definecolor{cyan}{rgb}{0,0.9,0.9}
\definecolor{orange}{rgb}{0.9,0.5,0}
\definecolor{magenta}{rgb}{1,0,1}
\definecolor{purple}{rgb}{0.8,0.4,0.8}
\definecolor{darkgreen}{rgb}{0.0,0.5,0.0}
\definecolor{gray}{rgb}{0.8242,0.8242,0.8242}
\definecolor{cadmiumgreen}{rgb}{0.0, 0.42, 0.24}
\definecolor{olive}{rgb}{0.5, 0.5, 0.0}

\newcommand{\reftab}[1]{{Table~\ref{#1}}}
\newcommand{\refsec}[1]{{Section~\ref{#1}}}
\newcommand{\reffig}[1]{{Figure~\ref{#1}}}
\newcommand{\refapp}[1]{{Appendix~\ref{#1}}}
\newcommand{\refeq}[1]{{Eq.~(\ref{#1})}}

\title[r-process nebula]{Modeling the emission lines from r-process elements in Supernova nebulae}
\author[G. Ricigliano et al.]{Giacomo Ricigliano$^{1}$\thanks{Contact e-mail: giacomo.ricigliano@gmail.com},
Kenta Hotokezaka$^{2}$
and Almudena Arcones$^{1,3,4}$
\\
$^{1}$Institut für Kernphysik, Technische Universität Darmstadt, Schlossgartenstr. 2, Darmstadt 64289, Germany\\
$^{2}$Research Center for the Early Universe, University of Tokyo, Hongo 7-3-1, Bunkyo, Tokyo 113-0033, Japan\\
$^{3}$GSI Helmholtzzentrum für Schwerionenforschung GmbH, Planckstr. 1, Darmstadt 64291, Germany\\
$^{4}$Max-Planck-Institut für Kernphysik, Saupfercheckweg 1, Heidelberg 69117, Germany}

\date{Accepted XXX. Received YYY; in original form ZZZ}

\pubyear{2025}

\begin{document}
\label{firstpage}
\pagerange{\pageref{firstpage}--\pageref{lastpage}}
\maketitle

\begin{abstract}
The origin of heavy r-process elements in the universe is still a matter of great debate, with a confirmed scenario being \ac{NS} mergers.
Additional relevant sites could be specific classes of events, such as \ac{GRB} \acp{SN}, where a central engine could push neutron-rich material outwards, contributing to the ejecta of the massive exploding star.
Here, we investigate our ability to infer the production of heavy elements in such scenarios, on the basis of the observed nebular emission.
We solve the steady-state ionization, level population, and thermal balance, for optically thin ejecta in \ac{NLTE}, in order to explore the role of heavy elements in cooling the gas, and their imprint in the emergent spectrum a few hundreds days post-explosion.
We find that heavy elements would be relevant in the cooling process of the nebula only if they account for at least $\sim1\%$ of the total ejected mass, at the typical kinetic temperatures of a few thousands K.
However, even in the absence of such amount, a few $0.1\%$ of the total ejected mass could be instead sufficient to leave a detectable imprint around $\sim1-10~{\rm\mu m}$.
This wavelength range, which would be relatively clean from features due to light elements, would be instead robustly populated by lines from heavy elements arising from forbidden transitions in their atomic fine structures.
Hence, the new generation of telescopes, represented by the \ac{JWST}, will most likely allow for their detection.
\end{abstract}
\acresetall


\section{Introduction}

The search for astrophysical sites capable of harboring r-process nucleosynthesis has been ongoing for more than half a century \citep{Burbidge:1957,Cameron:1957,Horowitz:2019,Cowan:2019pkx,Perego:2021dpw}.
Only recently, the detection of the \ac{GW} signal GW170817 \citep{TheLIGOScientific:2017qsa}, in coincidence with the optical to \ac{IR} \ac{EM} counterpart AT 2017gfo \citep{LIGOScientific:2017ync}, confirmed that \ac{BNS} mergers eject to some extent neutron-rich material undergoing r-process \citep{Lattimer:1974slx,Eichler:1989ve,Freiburghaus:1999}, the radioactivity of which powers the transient known as \ac{KN} \citep{Li:1998bw,Metzger:2010sy,Barnes:2013wka,Tanaka:2013,Metzger:2019zeh}.
Observations of radioactive isotopes on Earth \citep{Wallner:2015,Hotokezaka:2015zea}, as well as metal-poor stars in the Milky Way Halo and in dwarf galaxies such as Reticulum II \citep{Ji:2015wzg,Macias:2016sqg,Kirby:2020}, require the candidate r-process scenario to be rare, and high-yielding.
While these conditions can in principle be satisfied by \ac{BNS} mergers \citep[see, e.g.,][]{Hotokezaka:2018aui}, other types of events could still result in similar production rates.
Therefore, whether \ac{BNS} mergers account for the entirety of the observed r-process elements or not remains a matter of debate \citep{Hotokezaka:2018aui,Cote:2018qku,Molero:2020,Kobayashi:2022qlk,Beniamini:2023szg,Maoz:2024gch}.

Other candidates, for which still no evidence for r-process ignition has been firmly obtained, include special classes of \acp{CCSN}, such as \acp{MRSN}, where the resulting \ac{PNS} has a high degree of rotation and magnetization \citep{Symbalisty:1985,Nishimura:2005nz,Winteler:2012hu,Barnes:2017hrw,Moesta2018ApJ,Shankar:2021obg,Reichert:2023}.
Another example are collapsars, in which the collapsing massive star core produces a \ac{BH}-disk system similar to the \ac{NS} merger case \citep{MacFadyen:1998vz,Pruet:2002ky,Kohri:2005tq,Surman:2005kf,Siegel:2019,Siegel:2021ptt}.
In particular, despite the higher rarity of collapsars with respect to \ac{BNS} mergers, \citet{Siegel:2019} argued that the large amount of ejecta from these events could be the explanation for most of the r-process elements observed in the Universe \citep[but see also][for r-process nucleosynthesis in collapsar simulations]{Miller2020ApJ,Just2022ApJ,Fujibayashi2023ApJ,Dean2024PhRvD}.
Notably, the remarkable mass of \ce{^{56}Ni} produced (about a few $0.1~M_\odot$), coupled with the powerful central engine activity, results in significantly brighter \acp{SN}, when compared to normal \acp{CCSN}.
High accretion rates onto the central \ac{BH}, or the interaction between the \ac{PNS} magnetic field and its surrounding envelopes, can support the launch of relativistic jets giving rise to a long \ac{GRB} \citep{Woosley:1993wj,Wheeler:1999bu,Bucciantini:2007hy,Metzger:2010pp}.
Intriguingly, the recent detection of the long \ac{GRB} 211211A \citep{Rastinejad:2022} and \ac{GRB} 230307A \citep{Levan:2023ssd}, in association to a \ac{KN}-like emission, has reignited the question on the progenitor of such events, while providing new examples for a link between \acp{GRB} and r-process nucleosynthesis.

\Ac{SN} emissions observed in coincidence of a long \ac{GRB} have been characterized by very broad spectral lines at early times, corresponding to ejecta velocities up to $\sim0.1~c$ \citep[see, e.g.,][]{Modjaz:2015cca}.
The general absence of optical He I lines, in addition to H lines, classifies such events as \ac{BL} Type Ic \acp{SN}.
This indicates that the progenitor must have lost the H and most of the He envelope at some point during its evolution (\ac{SE} \ac{SN}), falling under the \ac{WR} class of stars \citep{Georgy:2009dd,Dessart:2011ns,Liu:2015gma}.
In this regard, the long \ac{GRB} 980425/\ac{SN} 1998bw represents a famous example of detection displaying the above features \citep{Galama:1998ea,Iwamoto:1998tg,Woosley:1999}.
With an estimated \ce{^{56}Ni} yield between $\sim0.1-0.5~M_\odot$, this event was likely attributed to the aspherical explosion of a rapidly rotating massive \ac{WR} star evolving at low metallicity \citep{Dessart:2017}.
While so far there have been several tens of \ac{GRB}/\ac{SN} associations, only \ac{SN} 1998bw and other two successive events, i.e. \ac{GRB} 060218/\ac{SN} 2006aj \citep{Mazzali:2006tk} and \ac{GRB} 171205A/\ac{SN} 2017iuk \citep{Wang:2018wqx}, where found at a distance closer than $\sim200$ Mpc, allowing for a more detailed study of the emission components.
Initially, analysis have been mainly focused on the optical wavelengths, both at early and late times, more than the \ac{IR} range, mostly due to the lower \ac{SNR} in that region.
However, the progressive upgrades in the capabilities of observatories such as the \ac{VLT}, and the launch of new space telescopes, like Spitzer or, now, the \ac{JWST}, has opened the possibility for a much more sensitive and deeper search \citep[see, e.g.,][]{Stritzinger:2023dea,Cartier:2024usw}.

In this context, to date there have not been investigations on the possibility of identifying r-process elements in the spectra of \ac{SE} \acp{SN}.
The first days of emission, corresponding to the photospheric phase of the ejecta, are generally characterized by an absorption spectrum poor in recognizable features, due to the high velocities blending the dense line population.
As time evolves and the ejecta enters the nebular phase, the spectrum morphs to be dominated by emission features.
In particular, heavy elements would be expected to contribute with lines in the \ac{IR} region, due to the lower opacity associated to the forbidden transitions between levels in their complex f-shell structures.
Following GW170817, renovated efforts have been spent in modeling accurately the nebular emission resulting from the \ac{NS} merger ejecta after $\sim1$ week of expansion \citep{Hotokezaka:2021,Hotokezaka:2022,Pognan:2022,Pognan:2022pix,Hotokezaka:2023aiq,Pognan:2023qhw,Pognan:2024ise}.
Such works resulted in the discovery of possible candidates in Se or W \citep{Hotokezaka:2022}, Te \citep{Hotokezaka:2023aiq}, and Rb \citep{Pognan:2023qhw}, in addition to those inferred from the photospheric phase, i.e. Sr or He \citep{Watson:2019xjv,Domoto:2021xfq,Gillanders:2022,Perego:2022,Tarumi:2023,Sneppen:2024,Sneppen:2024b}, Y \citep{Sneppen:2023}, and La and Ce \citep{Domoto:2022}.
These models have to take into account that the outflow is in \ac{NLTE}.
As such, they require the treatment of both radiative and collisional processes relevant in determining the atomic ionization and level population, from which the emission spectrum is derived.

In this work, we adopt and expand the nebular emission modeling proposed by \citet{Hotokezaka:2021}, in order to explore for the first time the possible presence of heavy elements in the ejecta from \ac{BL} Type Ic \acp{SN}.
The paper is structured as follows.
In \refsec{sec:model}, we describe the model employed, including its assumptions, and the treatment of the relevant atomic processes.
The input data required by the latter are listed in \refsec{sec:atomic_data} (integrated by \refapp{app:data}), while an exploration of its parameter space is conducted in \refsec{sec:model_applicability}.
We spend \refsec{sec:heating_role} and \refsec{sec:cooling_role} to note the overall impact that an r-process composition would have in the heating and cooling processes of the ejecta, respectively.
Our model is then benchmarked against a characteristic spectrum of a \ac{BL} Type Ic \ac{SN}, i.e. \ac{SN} 1998bw, in \refsec{sec:reproducing_sn1998bw}.
Finally, we investigate possible signatures or r-process elements in such a scenario in \refsec{sec:r-process_signatures}, discussing the robustness of our results in \refsec{sec:discussion}.
Our main takeaways are thus summarized in \refsec{sec:conclusion}.

\section{Nebula modelling}
\label{sec:model}

Following \citet{Hotokezaka:2021}, we model the nebular emission of an expanding ejecta in an explosive event by introducing a series of simplifying conditions.
First, we assume that the fluid is in a steady-state regime, i.e. that the atomic processes at the source of the emission are much faster than the fluid expansion.
This is reasonably valid at an early enough stage, where the recombination timescale is much shorter than the dynamical timescale \citep{Pognan:2022}.
This condition holds in our model, where the typical electron number density is $\gtrsim 10^{6}~{\rm cm^{-3}}$ at $\mathcal{O}(100)$ days after the explosion.
Second, we consider the ejecta to be optically thin, whereas the final spectrum will be dominated by pure emission.
Such an assumption heavily simplifies our photon transport modelling, which is limited to the effective treatment of the local self-absorption.
However, we recall that this approach loses validity when different lines contribute in the same wavelength region, as it can be the case for optical emissions from lanthanides.

We describe the ejecta at a fixed epoch using a multi-zone approach, where each zone is characterized by a density value and an elemental composition.
In order to determine the ion abundances within a zone, we solve the steady-state ionization balance:
\begin{equation}
    \label{eq:ion_balance}
    \frac{{\rm d}n_i^p}{{\rm d}t} = - \gamma^p_{i,{\rm tot}} n_i^p + \alpha_{i+1}^p n_{i+1}^p n_{\rm e} \approx 0 \, ,
\end{equation}
where $n_i^p$ indicates the number density of an element $p$ in ionization stage $i$ (hereafter species $\{p,i\}$), such that $\sum\limits_{p,i}n_i^p=n$ and $\sum\limits_{p,i}in_i^p=n_{\rm e}$, with $n$ and $n_{\rm e}$ the atoms and free electrons number density respectively.
Here, $\alpha_{i+1}^p$ is the total recombination rate coefficient for the species $\{p,i+1\}$ to recombine into $\{p,i\}$, obtained in general by summing the contributions from both radiative and dielectronic recombination processes.
Concurrently,
\begin{equation}
    \label{eq:ion_rate}
    \gamma^p_{i,{\rm tot}} = \gamma^p_{i,{\rm rad}} + \gamma^p_{i,{\rm rec}}
\end{equation}
is the total ionization rate per atom for the species $\{p,i\}$ to ionize into $\{p,i+1\}$.
\refeq{eq:ion_rate} shows the relevant contributions to ionization.

At the times of the nebular phase, we assume that the main source of ionization is represented by the radioactivity of the ejected material.
Radioactive energy is mainly released via $\beta$-particles produced in $\beta$-decays and $\gamma$-rays produced in nuclear de-excitations, with the second channel typically prone to leak out of  the ejecta due to the low medium opacity, leaving the first channel responsible for the energy deposition.
The heating rate as a function of time depends on which decaying species dominate the ejecta composition, ranging from a simple exponential chain, as in a Ni-powered \ac{SN}, to a power-law, as in the case of an r-process-powered \ac{KN}.
However, the energy deposition rate soon deviates from the instantaneous radioactive energy generation, as the thermalization of $\beta$-particles becomes inefficient.
For arbitrary r-process compositions, we treat radioactivity and thermalization using the model described in \citet{Hotokezaka:2020}.
Given a thermalized heating rate per unit volume $\Gamma$, the ionization rate per atom due to radioactivity is then
\begin{equation}
    \gamma^p_{i,{\rm rad}} = \frac{\Gamma}{w^p_i n^p_i} \, ,    
\end{equation}
where $w^p_i$ is the work per ion pair, i.e. the energy dissipated by a produced $\beta$-particle in causing one $\{p,i\}$ ionization event in the stopping plasma.
Here, we describe the work per ion pair as $w^p_i = w I^p_i$, with $I^p_i$ the species $\{p,i\}$ first ionization potential, and $w$ a model parameter.
Informed by the estimates of \citet{Hotokezaka:2021}, we fix the value of $w$ to $15$ throughout this work.

Secondly, we include in our model the contribution to ionization due to recombination photons, for which the ejecta is optically thick in the nebular phase.
This effect can be particularly important at low densities, where lightly-ionized species are favoured.
We define the probability for a photon created in the recombination event of species $\{q,j\}$ to ionize species $\{p,i\}$ as
\begin{equation}
    P^{pq}_{ij} = \int (1 - e^{-\tau}) \frac{n^p_i \sigma^p_i}{\sum\limits_{r,k} n^r_k \sigma^r_k} \left( \frac{{\rm d}N_{\rm ph}}{{\rm d}\nu} \right)^q_j {\rm d}\nu \, ,
\end{equation}
where $\sigma^p_i$ is the photoionization cross-section of species $\{p,i\}$, $\left( \frac{{\rm d}N_{\rm ph}}{{\rm d}\nu} \right)^q_j$ is the spectrum of photons emerging from the recombination of species $\{q,j\}$, and $\tau$ is the photon optical depth, which we define as
\begin{equation}
    \tau = R \sum_{r,k} n^r_k \sigma^r_k \, , 
\end{equation}
with $R$ the dimension of the zone.
While photons emerging from radiative recombinations have roughly energies close to the first ionization potential of the recombining species, the photons cascade of dielectronic processes has a generally unknown spectrum.
Here, we assume that photons are emitted in similar number at all energy scales between the first ionization potential of the recombining species and the first ionization potential of its neutral counterpart, i.e.
\begin{equation}
    \left( \frac{{\rm d}N_{\rm ph}}{{\rm d}\nu} \right)^q_j = \frac{1}{1 - \frac{I^q_0}{I^q_j + k_{\rm B}T_{\rm e}}} \frac{1}{\nu} \quad {\rm with} \quad I^q_0 < \nu < I^q_j + k_{\rm B}T_{\rm e} \, ,
\end{equation}
where we also account for a thermal energy budget.
The ionization rate per atom due to recombination photons is then
\begin{equation}
    \label{eq:ion_rate_rec}
    \begin{aligned}
        \gamma^p_{i,{\rm rec}} = & \sum_{q,j: I_j^q + k_{\rm B} T_{\rm e} \geq I_i^p} P^{pq}_{ij} \alpha_{j+1}^q n_{j+1}^q n_{\rm e} +\\
        & + \sum_{q,j: I_j^q + k_{\rm B} \bar{T}_{\rm e} \geq I_i^p} \bar{P}^{pq}_{ij} \bar{\alpha}_{j+1}^q \bar{n}_{j+1}^q \bar{n}_{\rm e} e^{-\bar{\tau}} \, ,
    \end{aligned}
\end{equation}
where the sum includes all recombining species able to produce photons energetic enough to ionize species $\{p,i\}$.
In \refeq{eq:ion_rate_rec}, we account for the contributions to ionization from recombination events taking place in the adjacent inner zone, indicated by the barred version of each quantity.
Such contributions are weighted with a diffusion factor depending on the optical depth of the inner zone.

The level population of each ion species is found by solving the steady-state level balance:
\begin{equation}
    \label{eq:lvl_pop}
    \begin{aligned}
        \frac{{\rm d}n_m}{{\rm d}t} = & \sum_{l: l \neq m} (n_l n_{\rm e} k_{lm} - n_m n_{\rm e} k_{ml}) +\\
        & - \sum_{l: l < m} \langle\beta_{ml}\rangle n_m A_{ml} + \sum_{l: l > m} \langle\beta_{lm}\rangle n_l A_{lm} \approx 0 \, ,
    \end{aligned}
\end{equation}
with $n_m$ the number density of an ion species $\{p,i\}$ in a level $m$, such that $\sum\limits_m n_m = n^p_i$.
Note that we dropped both the $p$ and $i$ indices for convenience, since the level equations for different species are completely decoupled, as a consequence of assuming the ejecta to be optically thin (apart from self-absorption).
\refeq{eq:lvl_pop} accounts for all electron collisional transitions between a level $m$ and any other level $l$, described by the collisional rate coefficient $k_{ml}$, as well as for all radiative transitions from a level $m$ to a lower level $l$, described by the radiative transition rate per atom $A_{ml}$, and characterized by a transition frequency $\nu_{ml}$.
We treat photon self-absorption by applying an escape probability $\langle \beta_{ml}\rangle$ to the radiative transition between state $m$ and $l$, which we define as
\begin{equation}
    \langle\beta_{ml}\rangle = \frac{1 - e^{-\tau_{ml}}}{\tau_{ml}} \, ,
\end{equation}
where $\tau_{ml}$ is the Sobolev optical depth for that transition.
For the collisional transitions, the collisional rate coefficient for de-excitations takes the form
\begin{equation}
    k_{lm} = \frac{8.63 \times 10^{-6} \Omega_{lm}(T_{\rm e})}{g_l T_{\rm e,1}^{\frac{1}{2}}} ~ {\rm cm^3~s^{-1}} \, ,
\end{equation}
with $l>m$, $g_l$ the level $l$ degeneracy, $\Omega_{lm}$ the correspondent collision strength, and $T_{\rm e,1}$ the kinetic temperature in units of ${\rm K}$.
Accordingly, the collisional rate coefficient for excitations follows
\begin{equation}
    k_{ml} = \frac{g_l}{g_m} k_{lm} e^{-\frac{E_{lm}}{k_{\rm B}T_{\rm e}}} \, .
\end{equation}

The contribution to the ejecta cooling rate density $\Lambda$ from a species $\{p,i\}$ is then computed as the sum over all the associated radiative de-excitations:
\begin{equation}
    \Lambda = \sum_{m,l:m>l} n_m E_{ml} \langle\beta_{ml}\rangle A_{ml} \, ,
\end{equation}
where $E_{ml}$ is the energy gap between level $m$ and $l$.
The ion and level populations are thus solved iteratively for different values of the kinetic temperature $T_{\rm e}$, until thermal balance is satisfied, i.e.
\begin{equation}
    \Gamma - \sum_{p,i} \Lambda^p_i \approx 0 \, ,
\end{equation}
where we reintroduced the indices $p$ and $i$ in order to indicate the different ions.

From the above ejecta snapshot, we derive an emission spectrum by summing over all the lines of every ion.
The contribution to the luminosity from a single ion species $\{p,i\}$ is then
\begin{equation}
    \label{eq:spectrum}
    \begingroup
    \mathcode`v=\varv
    L_\nu = \sum_{m,l:m>l} N_m E_{ml} \langle\beta_{ml}\rangle A_{ml} \mathcal{N}_\nu\left(\nu_{ml},\left(\frac{\nu_{ml}v}{\sqrt{2}c}\right)^2\right) \, ,
    \endgroup
\end{equation}
where $N_m$ is the number of atoms in level $m$, while $\mathcal{N}_\nu$ is the normal distribution, which we use to characterize the line spread due to the zone velocity $\begingroup \mathcode`v=\varv v \endgroup$ shift.
Note that in the last equation we dropped again the $p$ and $i$ indices for simplicity.
We reintroduce them right after, as, finally, the total luminosity is thus
\begin{equation}
    L_\nu = \sum_{p,i} L^p_{i,\nu} \, .
\end{equation}

\section{Atomic data}
\label{sec:atomic_data}

We report here the data used to solve the ionization balance of the ions and model the nebular emission.
The remaining set of atomic data necessary to calculate the level population and the emergent spectrum, including energy levels, radiative transition rates, and collision strengths, is listed in \refapp{app:data}.

\subsection{M1 line list for r-process elements}

We use the \ac{M1} line list provided by \cite{Hotokezaka:2022}.
This list includes \ac{M1} lines of heavy elements with experimentally calibrated energy levels \citep{NIST_ASD}.
The radiative transition rates are obtained by using an analytic formula based on LS coupling (see \refapp{app:data}).

\subsection{Recombination rates}

We import radiative and dielectronic recombination rates from the available literature and public databases, and we integrate them with our previously-computed data.
We include ions up to the fourth ionization stage, as we expect a mild level of ionization at the kinetic temperatures considered $\mathcal{O}(10^3~{\rm K})$.
For the recombination of Fe I-III, and Ni II, we use the values computed using the R-matrix method in an ab-initio framework by \citet{Nahar:1996,Nahar:1997a,Nahar:1997b} and \citet{Nahar:2001} respectively.
Because of the absence of data consistent with the works above, for Ni I, Ni III, and Co ions we assume the values of the respective Fe ionization stages.
For lighter elements, we use the analytic fits for both radiative and dielectronic recombination rates provided by \citet{Badnell:2006,Bleda:2022} and references therein.
Such fits are based on autoionization transition rates computed using the Breit-Pauli Hamiltonian with the \texttt{AUTOSTRUCTURE} codes suite \citep{Badnell:1986}, and cover ions up to the P plus Ar isoelectronic sequences.
Therefore, in order to fill in the missing data, we integrate the above rates with those provided by the CHIANTI atomic database \citep{Dere:1997,Zanna:2021}, with the caveat that at low electron temperatures $\lesssim\mathcal{O}(10^2~{\rm K})$, the dielectronic contribution to the total recombination rate is not always available.
Finally, for elements heavier than Ni, we assume the same dielectronic recombination rates as were obtained by \citet{Hotokezaka:2021} for Nd\footnote{The included rates describe Nd recombinations from stage IV to III, and III to II. We use the latter also for stage II to I.}, using the atomic structure code \texttt{HULLAC} \citep{Bar-Shalom:2001}, and we use the formula provided by \citet{Axelrod:1980} for the radiative recombination of Fe ions:
\begin{equation}
    \alpha_{\rm RR} = 3 \times 10^{-13} i^2 \left[\left(\frac{T_{\rm e}}{10^4~{\rm K}}\right)^{-\frac{3}{2}} - \frac{1}{3}\left(\frac{T_{\rm e}}{10^4~{\rm K}}\right)^{-\frac{1}{2}}\right] ~ {\rm cm^3~s^{-1}} \, ,
\end{equation}
with $i$ the recombined ion charge.

In this regard, we note that current efforts in producing detailed atomic data for \acp{KN} have recently resulted in new total recombination rates for light r-process elements, including the first three ionization stages of Se, Rb, Sr, Y, and Zr \citep{Banerjee:2025eoh}.
The latter demonstrated how accurate rate calculations can substantially change the predictions on the observed nebular spectra.
Despite here we do not include such data yet, leaving it to the further development of our model, we explore the effects of different ionization levels on our predictions in \refsec{sec:discussion}.

\subsection{Photoionization cross-sections}

We use the photoionization cross-sections for Fe ions computed with \texttt{HULLAC} by \citet{Hotokezaka:2021}, in order to characterize all non-neutral species.
Photoionization on neutrals is instead described starting from the photoabsorption calculations stored in the XCOM database of the \ac{NIST} \citep{Scofield:1973}.
Since the latter consider incident photon energies ranging from 1 to 1500 keV, thus covering only the high energy tail of the recombination photons' distribution, we assume a flat cross-section for lower energies, in continuity with the high energy part.

\section{Model applicability}
\label{sec:model_applicability}

\begin{figure*}
    \centering
    \includegraphics[width=\textwidth]{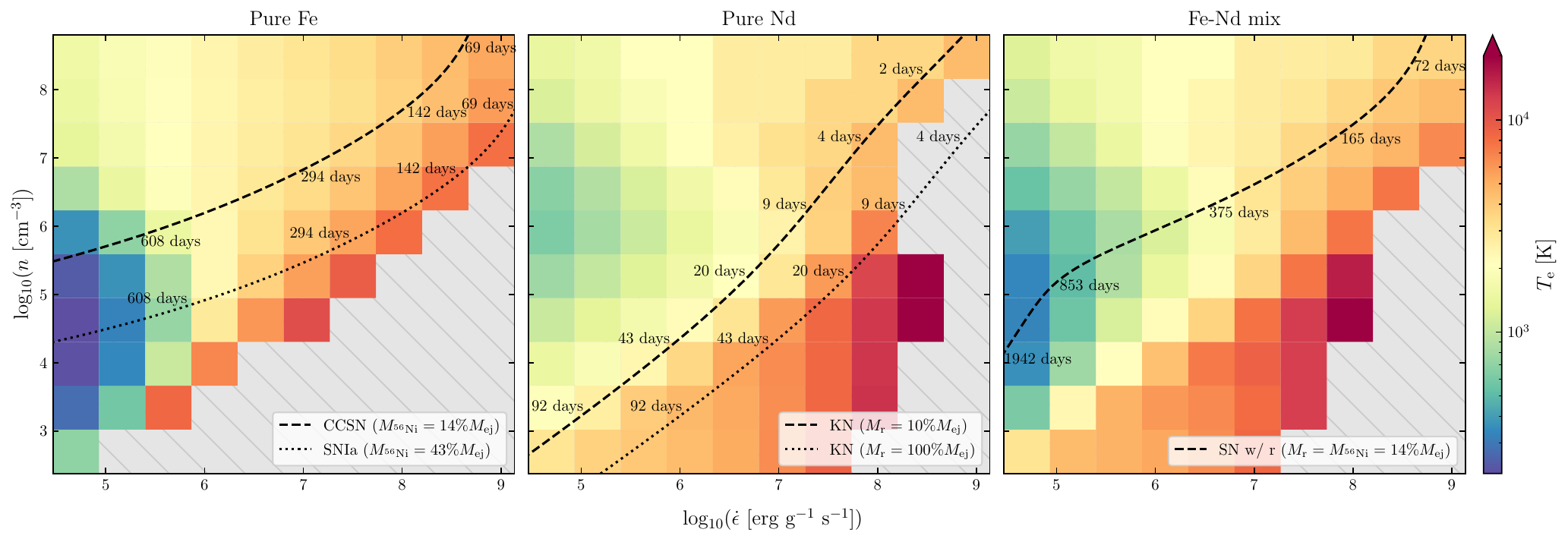}
    \caption{Electron temperature as a function of the specific energy deposition rate and the number density of atoms, for a gas of ions (I-IV) of pure Fe (left panel), pure Nd (central panel), and a mixture of $50\%$ Fe and $50\%$ Nd in mass fraction (right panel).
    Values are computed on a 10x10 grid log-spaced, using ionization and thermal balance.
    For approximate visual reference, we superimpose plausible trajectories with their time evolution of a \ac{CCSN}, a \ac{SN} Ia, a \ac{KN} with two degrees of r-process material dilution, and a \ac{CCSN} polluted with r-process material.}
    \label{fig:grid_Te}
\end{figure*}

\begin{figure*}
    \centering
    \includegraphics[width=\textwidth]{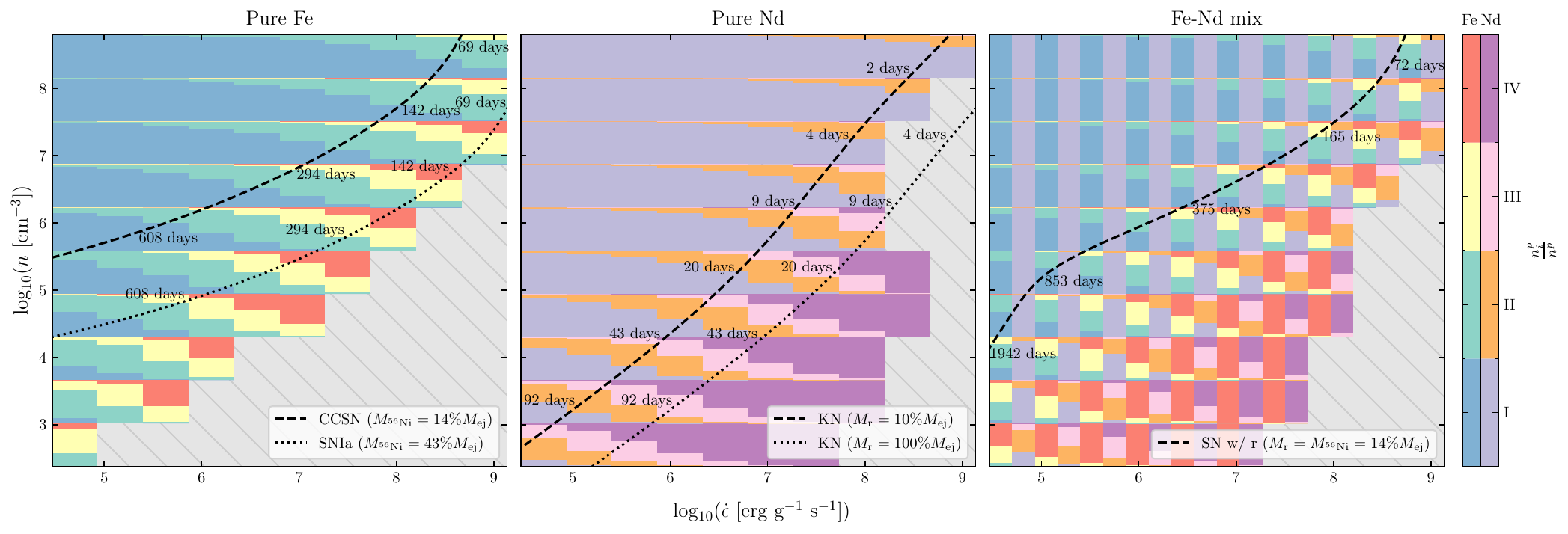}
    \caption{Same as \reffig{fig:grid_Te}, with the plotted quantity being the number fraction of ions as stacked histogram.
    For the Fe-Nd mixture case (right panel), the ion fractions of Fe and Nd are displayed on the left and the right side of each grid cell, respectively.}
    \label{fig:grid_frac}
\end{figure*}

The absence of a complete database including all possible allowed and forbidden atomic transitions for every element inherently hinders the computation of a fully accurate spectrum emergent from a heterogeneous composition consisting of both light and heavy elements.
However, using the model described in \refsec{sec:model} together with the data listed in \refsec{sec:atomic_data}, we can already obtain useful estimates, assuming the validity of all the assumptions and approximations employed.
In order to test our setup, we define a one-zone model parameter space which should cover the typical conditions found in the nebular phase of expanding ejecta.
For a given composition, we identify two main degrees of freedom for our model, i.e. the amount of heating injected in the system per unit time and unit mass $\dot{\epsilon}$, and the number density of atoms $n$.
We then focus on relevant intervals for these quantities, as suggested by the typical evolution of ejecta from different kind of events.
For this purpose, we construct plausible trajectories for different \acp{SN} and \acp{KN}.
Despite in this work we are not specifically interested in each one of these events, we take the opportunity to test the performance of our model in all of the above cases.
For the ejecta density, we consider a time-dependent homologous expansion law of the form
\begin{equation}
    n(t) = n_0 \left(\frac{t}{t_0}\right)^{-3} \, ,
\end{equation}
where we arbitrarily set an initial time $t_0=10^4$ s, and the corresponding initial density is computed as average over a sphere with mass $M_{\rm ej}$ and velocity $\begingroup \mathcode`v=\varv v_{\rm ej} \endgroup$ depending on the case.
We thus model a \ac{CCSN} with a typical ejecta mass and velocity of $M_{\rm ej}^{\rm CCSN}=0.5~M_\odot$ and $\begingroup \mathcode`v=\varv v_{\rm ej}^{\rm CCSN}=0.01~c \endgroup$ respectively, and we compare it to a \ac{SN} Type Ia with $M_{\rm ej}^{\rm SNIa}=1.4~M_\odot$ and $\begingroup \mathcode`v=\varv v_{\rm ej}^{\rm SNIa}=0.03~c \endgroup$.
Additionally, we construct a \ac{KN} with $M_{\rm ej}^{\rm KN}=10^{-2}~M_\odot$ and $\begingroup \mathcode`v=\varv v_{\rm ej}^{\rm KN}=0.1~c \endgroup$.
The heating evolution counterpart is computed from the \ce{^{56}Ni-Co} and \ce{^{44}Ti-Sc} decay chains energy input in the \ac{SN} cases, while for the \ac{KN} we assume the characteristic r-process heating rate from \citet{Metzger:2019zeh}:
\begin{equation}
    \dot{\epsilon}(t) = 2 \times 10^{10} \left(\frac{t}{1~{\rm day}}\right)^{-1.3} \, .
\end{equation}
In the \ac{SN} cases we consider a \ce{^{56}Ni} mass of $M_{\rm \ce{^{56}Ni}}^{\rm CCSN}=6.9\times 10^{-2}~M_\odot$ (the value inferred for SN1987A) and $M_{\rm \ce{^{56}Ni}}^{\rm SNIa}=0.6~M_\odot$, for the \ac{CCSN} and the \ac{SN} Type Ia models respectively.
From the \ac{KN} model instead, we differentiate two cases, by assuming an r-process mass of $M_{\rm r}^{\rm KN}=10^{-3}~M_\odot$ and $M_{\rm r}^{\rm KN}=10^{-2}~M_\odot$, respectively.
We also produce an alternative trajectory, by considering the \ac{CCSN} model, and by polluting it with an amount of r-process material equal to its amount of \ce{^{56}Ni}.
Finally, we compute the energy deposition rates by applying the following thermalization prescriptions.
In order to account for the thermalization of $\gamma$-rays, we employ the analytic formula from \citet{Sharon:2020}, specifically tailored on core-collapse \acp{SN}, but demonstrated to be a good approximation also for Type Ia \acp{SN} and \acp{KN} \citep{Guttman:2024}:
\begin{equation}
    f_{\gamma,{\rm th}}(t) = \frac{1}{(1 + \frac{t}{t_0}^n)^{\frac{2}{n}}} \, ,
\end{equation}
where $f_{\gamma,{\rm th}}(t)$ is the fraction of $\gamma$-ray energy instantaneously deposited in the ejecta, $t_0\sim100$ days is the characteristic $\gamma$-ray escape time, and $n\sim3$ is a parameter that controls the transition between optically thick and optically thin regime.
In parallel, we resort to the analytic formula derived by \citet{Barnes:2016umi} in order to characterize the energy deposition of $\beta$-particles from the decay of the r-process material, where we assume that $20\%$ of the total decay energy is carried by such channel, while $45\%$ by $\gamma$-rays and the remaining is lost with neutrinos.
The formula is based on an estimate of the characteristic time at which the thermalization of $\beta$-particles becomes inefficient, and it is parametrized by the total ejecta mass and average velocity.
Despite most of the above masses and velocities being only broadly informed on the literature and still somewhat arbitrary, the resulting trajectories cover well the parameter space region where these various explosive events live.

As visible in \reffig{fig:grid_Te}, after $\sim100$ days the ejecta from various \acp{SN} are heated with $\lesssim10^9~{\rm erg~s^{-1}~g^{-1}}$, and within $\sim700$ days the heating rate reaches $\sim10^5~{\rm erg~s^{-1}~g^{-1}}$.
Simultaneously, at each time, the density is lower by about one order of magnitude for the faster ejecta of the \ac{SN} Ia, compared to the \ac{CCSN} case.
Such time interval is expected to bracket the conditions necessary for the steady-state nebular phase of \acp{SN}, as illustrated by \citet{Axelrod:1980}.
For a typical \ac{KN} instead, the same range of conditions are reached much earlier in the ejecta evolution, between a few days and $\sim100$ days, with a steeper density decrease due to the considerably faster medium.
Notably, as further investigated in \refsec{sec:heating_role}, polluting \ac{CCSN} ejecta with even a considerable amount of r-process material does not relevantly impact the heating before a few years, when the \ce{^{56}Co} decay fades to be soon replaced by the \ce{^{44}Sc} decay as main energy source.
However, already in the steady-state phase, r-process material may still have a role in the ejecta cooling mechanism (see \refsec{sec:cooling_role}).

Also shown in \reffig{fig:grid_Te} is the kinetic temperature behaviour within the parameter space considered, for different one-zone toy compositions.
The calculations are carried on a 10x10 log-spaced grid, coupling the ionization balance and the thermal balance to obtain both the correspondent ionic composition and temperature.
For simplicity, we consider a system composed by pure Fe (left panel), pure Nd (central panel), and a mixture of Fe and Nd present in equal mass.
We stress here that such toy compositions are not meant to be a precise proxy for the systems the trajectories of which have been described above.
The correct thermodynamic status of such systems can only be obtained by considering a full realistic time-dependent composition, including elements both lighter and heavier than Fe, several of them potentially having a relevant cooling role.
We address this point in \refsec{sec:cooling_role}.
Instead, here we focus on the considered elements (and a first direct combination of them) for illustrative purposes, grasping the typical behaviour around the conditions of astrophysical interest, whereas such behaviour is qualitatively expected also in richer compositions.

In general, the grid values reflect the cooling ability of the ions, which increases as a function of $T_{\rm e}$.
Given the imposed thermal balance, as expected $T_{\rm e}$ always monotonically increases with the input heating, for a fixed density.
Roughly all ejecta trajectories lie in a region where the mixture temperature is a few $10^3$ K in the steady-state phase, while regions with higher (lower) density and lower (higher) heating are not astrophysically explored.
Still, when looking at the different kinds of ejecta, the temperature for Fe shows a stronger excursion, decreasing during the \acp{SN} evolution, compared to Nd in \acp{KN}.
In the latter case, $T_{\rm e}$ maintains a rather constant value, such that, depending on the ejecta mass and velocity, it can also potentially re-increase, as found by \citet{Pognan:2022} using a much more sophisticated transport model.
The result of combining Fe and Nd in very optimistic equal amounts, is the product of a balance between opposing effects.
On one side, the amount of each element is trivially reduced, and, with it, its contribution to the overall cooling.
On the other side, the presence of different species helps relieving the burden of cooling from a single element, with the most efficient cooler dominating, depending on the specific conditions.
In our Fe-Nd case, the main effect of such combination is only a slight redistribution of the kinetic temperature in the grid which contributes to smooth out the sharper landscape of the pure Fe case.

On a different note, \reffig{fig:grid_Te} uncovers a model limit, whereas the uncolored cells correspond to a region where the thermal balance calculation does not converge.
This is due to the inability of our model to cool enough in order to compensate for the injected heating, and it is clearly dependent on the considered elemental composition.
While it is possible that in this region the system is physically unable to meet the thermal balance constraint within the model framework and assumptions, we note that this limit can also be affected by the specific line list included in the input data (see \refsec{sec:atomic_data}), which can substantially change the cooling function of each element.

Completing the picture described above, \reffig{fig:grid_frac} shows the number fraction of ion species for each of the toy models considered.
In all cases, a lower kinetic temperature of the system leads to a generally lower degree of ionization, with Fe stages I-II favoured over stages III-IV for the instantaneously denser \acp{CCSN}, compared to the \ac{SN} Ia case.
Similarly, in the \acp{KN}, a lower instantaneous heating due to less r-process material leads to the same effect for Nd ions.
This is caused by an increase in the recombination rates for lower temperatures, which are determined by less heating and thus also lower non-thermal ionization rates.
However, this trend is not universal, as evident from the far left region of the parameter space, where, especially for Fe, despite the very low temperatures, the stage II survives.
In fact, as listed in \refsec{sec:model}, the computed ionization balance is determined also by the contribution to ionization coming from recombination photons.
We asses the impact of such ionization component in \refapp{app:photoionization_impact}.
Finally, when mixing the two elements, we note that for a fixed kinetic temperature the average ionization stage of Fe is always greater than that of Nd, as a result of the lower recombination rates of Fe ions.

\section{Heating role}
\label{sec:heating_role}

\begin{figure}
    \centering
    \includegraphics[width=\columnwidth]{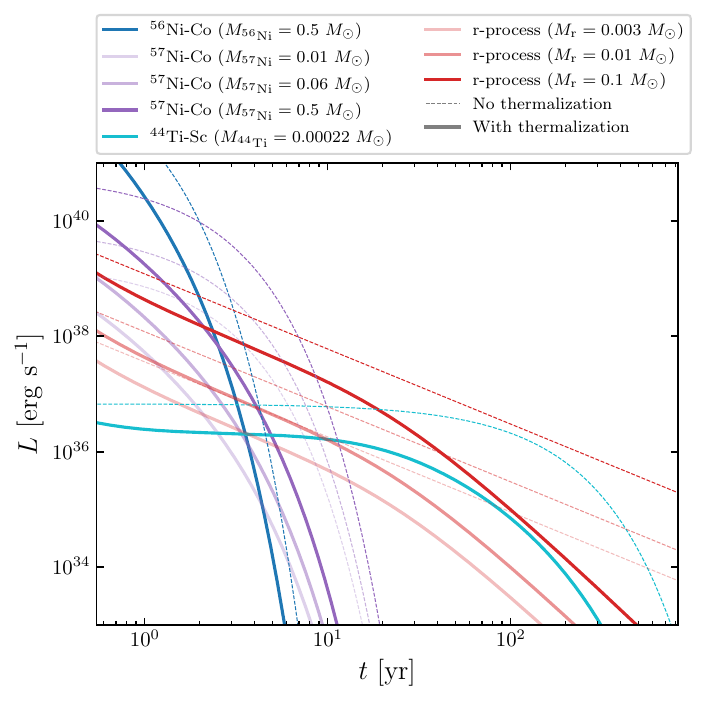}
    \caption{Total radioactive heating from the decay chains of \ce{^{56}Ni-Co} and \ce{^{44}Ti-Sc} for a \ac{SE} \ac{SN}, compared to the decay chain of \ce{^{57}Ni-Co} for a variable initial amount of \ce{^{57}Ni}, and to the heating due to the decay of a variable amount of r-process material.
    The values of \ce{^{56}Ni} and \ce{^{44}Ti} are taken from the CO138H \ac{SN} model by \citet{Nomoto:2000}.
    The radioactive energy produced (dashed lines) is shown for visual reference in contrast to the deposited energy (solid lines).}
    \label{fig:heating}
\end{figure}

In the context of \ac{SE} \acp{SN}, the importance that r-process elements may have in heating the ejecta is directly in competition with the heating from radioactive Ni.
Most likely, even an amount of heavy elements relevantly exceeding the typical estimates for \ac{BNS} mergers would not be sufficient to leave a signature in the \ac{SN} light curve near peak time, due to the much higher specific energy released by the massive \ce{^{56}Ni} content.
However, while the timescales considered in such case correspond to the photospheric phase of the emission, in principle the evolving astrophysical conditions and nuclei involved could lead to a different outcome during the nebular phase.

\reffig{fig:heating} shows a simple estimate of the total amount of instantaneous heating coming from the decay chain of \ce{^{56}Ni-Co} and \ce{^{44}Ti-Sc}, compared to the heating from typical r-process material in variable amount, in the case of a \ac{SE} \ac{SN}.
We also explore the presence of a significant initial amount of \ce{^{57}Ni}, the decay of which could also manifest in the light curve.
We consider the mass of \ce{^{56}Ni} produced in the explosion to be $M_{\ce{^{56}Ni}}=0.5~M_\odot$, and that of \ce{^{44}Ti} to be $M_{\ce{^{44}Ti}}=2.2\times10^{-4}~M_\odot$, as obtained by \citet{Nomoto:2000} in their CO138H \ac{SN} explosion model.
Such value of the \ce{^{56}Ni} mass roughly represents a conservative upper bound, as inferred from the light curves of \ac{SE} \acp{SN} analyzed by \cite{Rodriguez:2024}.
From the same CO138H model, we also take into account the amount of \ce{^{57}Ni}, i.e. $M_{\ce{^{57}Ni}}=1.5\times10^{-2}~M_\odot$, as a basis from which exploring its relevance.
For the thermalization of the decay energy, we use the prescriptions introduced in \refsec{sec:model_applicability}, assuming a total ejecta mass of $10~M_\odot$ and average velocity $0.02~c$ \citep[see, e.g.,][]{Nakamura:2001}.

As shown in \reffig{fig:heating}, a few times $10^{-3}~M_\odot$ of r-process material would be sufficient to eventually emerge in the light curve, in a limited time interval.
A mass of r-elements of a few times $10^{-2}~M_\odot$ would then dominate the emission for several years between the \ce{^{56}Co} and the \ce{^{44}Sc} contributions, potentially being comparable to the decay of \ce{^{44}Sc} around present time.
Higher amounts would completely eclipse the \ce{^{44}Ti} decay chain source, dominating the overall energy release and lead the emission already from a few years post-explosion on, once the \ce{^{56}Co} decay has past its activity peak.

On the other hand, the presence of r-process elements could be partially hidden also by the heating from the \ce{^{57}Ni} decay chain.
While the CO138H model from \citet{Nomoto:2000} predicts a mass of \ce{^{57}Ni} which is too low to be visible, starting from a relative amount of $\ce{^{57}Ni}/\ce{^{56}Ni}\sim0.12$, the imprint of \ce{^{57}Co} decay would be relevant in the same time window as the r-process heating.
However, because of the steeper decrease in the decay heating, even a very optimistic content of \ce{^{57}Ni} matching the one of its lighter isotope ($\ce{^{57}Ni}/\ce{^{56}Ni}\sim1$) would leave room for $\sim10^{-2}~M_\odot$ of r-process material to be seen.

Notably, these alternative sources of energy would reveal themselves in the observed luminosity, due to the different slope, and assuming no additional source as a central engine.
In the case of \ac{SN} 1998bw, however, the time interval of interest is currently left unexplored, since the most recent work reconstructing a quasi-bolometric luminosity from available photometric data includes observations only up to $\sim500$ days post-\ac{GRB} \citep{Clocchiatti:2011}.

\section{Cooling role}
\label{sec:cooling_role}

\begin{figure*}
    \centering
    \includegraphics[width=\textwidth]{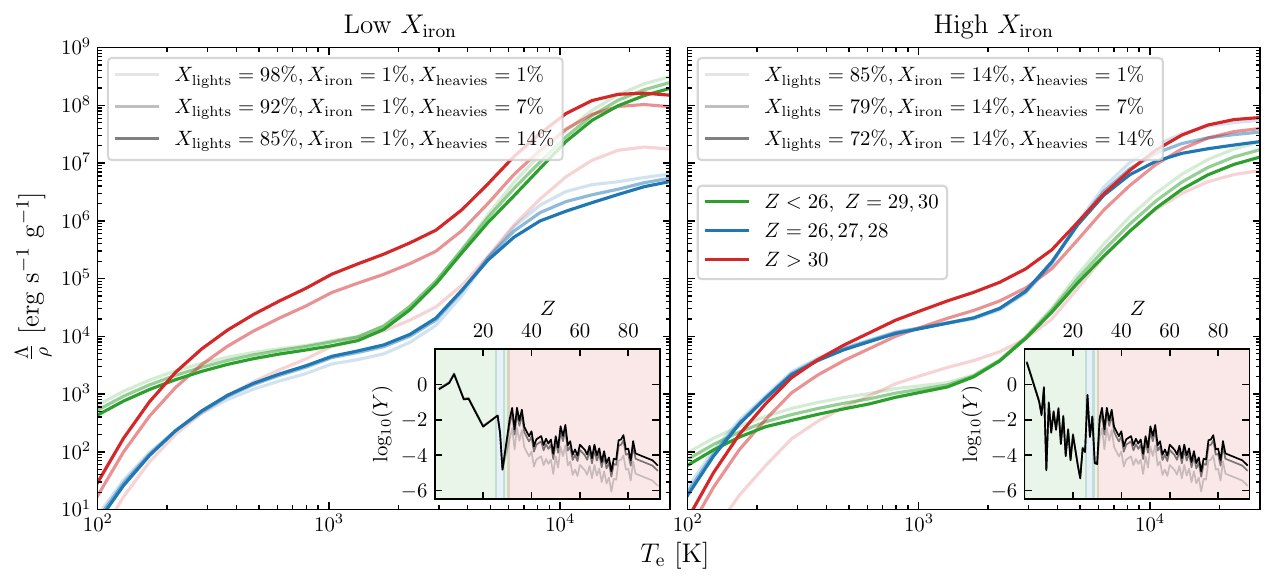}
    \caption{Contribution to the specific cooling coming from light elements ($Z<26$, $Z=29,30$), iron group elements ($Z=26,27,28$) and heavy elements ($Z>30$), for different mixtures of such groups.
    The specific cooling rate is shown as a function of the kinetic temperature at a fixed density of $10^6~{\rm cm^{-3}}$, and it is obtained through an iterative computation of the thermal (and ionization) balance.
    Two sets of variations are considered, i.e. a case with low amount of iron (left panel), and a case with high amount of iron (right panel).
    For each set, the amount of heavy elements is progressively increased at the expenses of light elements.
    Each panel also shows the set of correspondent abundance patterns, with the abundance of light elements taken from the S25 model computed by \citet{Rauscher:2001dw} (right panel), and from the r0e2 model computed by \citet{Dessart:2017} (left panel), while the abundance of heavy elements being the solar r-process pattern derived by \citet{Prantzos2020}.}
    \label{fig:cooling}
\end{figure*}

Independently from the role played in heating up the ejecta, a sufficient amount of heavy elements could add a relevant contribution to the cooling of the system already in the steady-state nebular phase, when the \ce{^{56}Co} decay is supposed to be the main source of energy.
We investigate the r-process material cooling power, by considering sets of different toy mixtures, constructed as follows:
\begin{itemize}
    \item for elements with $Z\leq30$, we use alternatively the abundance pattern of the S25 \ac{SN} model from \citet{Rauscher:2001dw}, based on a $25~M_\odot$ \ac{ZAMS} progenitor, or the r0e2 model from \citet{Dessart:2017}, based on a $40~M_\odot$ \ac{ZAMS} progenitor (see also \refsec{sec:nebular_spectra});
    \item for elements with $Z>30$, we use the solar r-process component pattern as derived by \citet{Prantzos2020}.
\end{itemize}
We thus group the elements into light ($Z\leq30$) and heavy ($Z>30$) ones, and we extract from the first group a third one constituted by the iron group elements with $Z=26,27,28$.
Their total mass fractions, i.e. $X_{\rm lights}$, $X_{\rm heavies}$, and $X_{\rm iron}$, respectively, are then varied in order to create the different sets of abundances.
We start by considering two compositions without heavy elements, representing considerably different base scenarios, with a higher relative production of iron coupled with the S25 model, and a lower relative production of iron combined with the r0e2 model.
These cases are generally inspired by a \ac{CCSN} such as \ac{SN} 1987A, and by a \ac{SE} \ac{SN} such as \ac{SN} 1998bw, respectively.
We note here that the mass fraction of iron does not reflect in general the absolute amount of iron produced, which is in fact higher in, e.g., \ac{SN} 1998bw, when compared to \ac{SN} 1987A.
Yet, the mass fraction is in this case the determinant factor when investigating the dominance of different species in terms of cooling power within a single zone mixture.
From these starting compositions, the fraction of heavy elements is progressively increased from $1\%$ to $7\%$, and finally to $14\%$ of the total ejecta mass, at the expenses of the light elements.
We consider $X_{\rm heavies}=14\%$ as a limiting case where the presence of heavy elements is ideally boosted to a rather unrealistic value, for illustrative purposes.
For each of the resulting 6 custom compositions, we build the total cooling function of the three elemental groups, by computing ionization and thermal balance over a range of kinetic temperatures and at a fixed density of $10^6~{\rm cm^{-3}}$.
The choice of fixing the density is well supported by the fact that the dependence of the cooling function on this parameter is generally minor \citep{Hotokezaka:2021}.
The thermal balance at each $T_{\rm e}$ is ensured through an iterative process that updates the value of the input heating rate in order to match the total cooling rate.

We show the variation in the different contributions to the cooling associated to the variation in the composition in \reffig{fig:cooling}, with the right panel showing the high $X_{\rm iron}$ case, and the left panel the low $X_{\rm iron}$ case.
In absence of heavy elements, iron tends to easily dominate the energy loss, primarily at temperatures from a few hundreds to $\sim10000$ K.
Even when the mass of the iron group elements is substantially lowered from $14\%$ to just $1\%$ of the total mass, the iron contribution is still only a factor of a few lower than that of the remaining elements combined, for $T_{\rm e}\sim1000~{\rm K}$.
The prominent role of Fe, Co and Ni lines in the nebular phase of \acp{SN} is a well established fact, as evidenced by the multitude of observed \ac{SN} spectra, and especially for Type Ia \acp{SN}, where the iron production is even greater.

Yet, in principle, also heavy elements could have a relevant impact, as already pointed out by \citet{Hotokezaka:2021} for the \ac{KN} case, whereas, for example, the cooling efficiency of Nd is more than one order of magnitude greater than that of Fe.
In fact, including a full composition, we observe similar cooling rates between heavy elements and iron across a wide range of temperatures, for these two groups having the same mass.
In particular, the cooling of heavy elements is more efficient by a factor of a few around $T_{\rm e}\sim1000~{\rm K}$, and can be substantially higher also for $T_{\rm e}\gtrsim10000~{\rm K}$.
In this regard, we note that, at those high temperatures, the cooling function of heavy species may continue to grow with $T_{\rm e}$.
However, we observe it saturating instead, likely because we do not model accurately all possible \ac{E1} and \ac{E2} transitions, taking over the \ac{M1} ones that instead dominate at lower temperatures.
Nevertheless, here we are interested only in the temperature region around a few thousands K, where the assumptions of our model are verified.
From our estimates, a few percent of the total mass in heavy elements is likely sufficient to contribute to the cooling similarly to the entire group of light species.
It follows that, depending on the amount of iron, a mass of heavy elements between $\sim1-10\%$ of the total mass is needed in order to observe their overall dominance at a few thousands K.
While the case with higher $X_{\rm iron}$ appears unlikely to harbor $\sim10\%$ of the total mass in heavy material, the scenario with lower $X_{\rm iron}$ could produce enough heavy species to, at least partially, become a relevant cooling channel for the ejecta.

We note that the above considerations are obtained by exploring different values for $X_{\rm lights}$, $X_{\rm heavies}$, and $X_{\rm iron}$, under the assumption that the specific choice of abundance patterns used to characterize the toy elemental mixtures is not crucial in this context.
In other words, here we assume that detailed variations in the pattern of the light or of the heavy elements, due to different astrophysical conditions as well as theoretical models, tend to roughly integrate out when computing their total contribution to the cooling rate, on a sufficiently large scale of temperatures.
In reality, in a similar fashion to what happens at the nuclear level in the case of the radioactive heating rate, we cannot exclude important contributions from distinguished atomic species to significantly alter the shape of the cooling functions.
However, we propose these ballpark estimates as a first step towards the recognition of the role of heavy elements in \acp{SN} nebulae.

\section{Nebular spectra}
\label{sec:nebular_spectra}

In addition to the cooling via r-process elements, we investigate their potential imprint on the subsequent spectrum, and its detectability.
For this purpose, we choose to target the well studied \ac{SN} 1998bw, as representative candidate of a scenario where heavy elements may be synthesized.

\subsection{Reproducing the late SN 1998bw emission}
\label{sec:reproducing_sn1998bw}

\begin{table}
    \centering
    \caption{List of relevant quantities characterizing each of the zones in our \ac{NLTE} model, used to target the \ac{SN} 1998bw spectra at 215.4 days post-explosion.
    The values are informed on the r0e2 explosion model from \citet{Dessart:2017}.
    From top to bottom, we list the zone total mass $M_{\rm tot}$, average velocity $\begingroup \mathcode`v=\varv v_{\rm avg} \endgroup$, atomic number density $n$, specific radioactive heating rate $\dot{\epsilon}$, and average atomic mass number $\Bar{A}$.
    We also report the kinetic temperature $T_{\rm e}$ and electron fraction $\chi_{\rm e}=\frac{n_{\rm e}}{n}$, as obtained by running the ionization and thermal balance in each zone.}
    \begin{tabular}{ll|ccc}
    \toprule
    Zone             &                                     & $1^\circ$ & $2^\circ$ & $3^\circ$ \\
    \midrule
    $M_{\rm tot}$    & [$M_\odot$]                         & 1.0       & 5.8       & 2.4       \\
    $\begingroup \mathcode`v=\varv v_{\rm avg} \endgroup$    & [$c$]                               & 0.007     & 0.020     & 0.030     \\
    $n$              & [$10^7~{\rm cm^{-3}}$]              & 4.88      & 5.20      & 0.19      \\
    $\dot{\epsilon}$ & [$10^7~{\rm erg~s^{-1}~g^{-1}}$]    & 2.56      & 0.26      & 0.26      \\
    $\Bar{A}$        &                                     & 23.1      & 17.3      & 14.7      \\
    \midrule
    $T_{\rm e}$      & [K]                                 & 4058      & 4861      & 5648      \\
    $\chi_{\rm e}$   &                                     & 0.26      & 0.07      & 0.30      \\
    \bottomrule
    \end{tabular}
    \label{tab:spectra_model}
\end{table}

\begin{figure}
    \centering
    \includegraphics[width=\columnwidth]{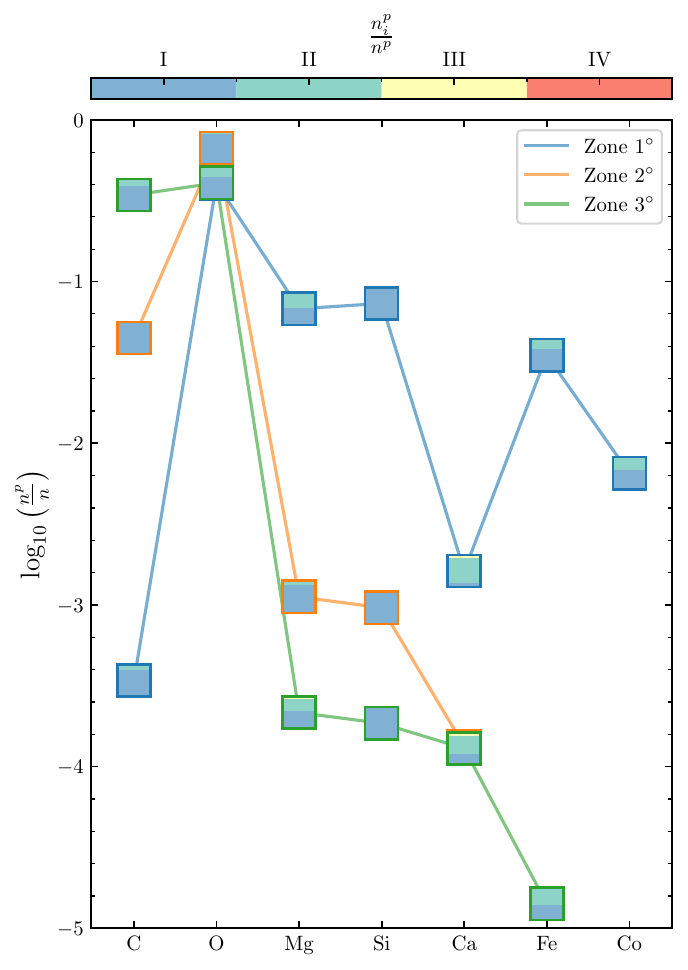}
    \caption{Atomic number fraction $\frac{n^p}{n}$ of the most abundant elements included in our \ac{NLTE} model, targeted on \ac{SN} 1998bw at 215.4 days, and informed on the r0e2 model from \citet{Dessart:2017}.
    The composition is shown for each of the three constructed zones, and the relative amount of the different ionization stages $\frac{n^p_i}{n^p}$, as obtained from ionization and thermal balance, is displayed for each element as a stacked histogram.}
    \label{fig:composition}
\end{figure}

\begin{figure*}
    \centering
    \includegraphics[width=\textwidth]{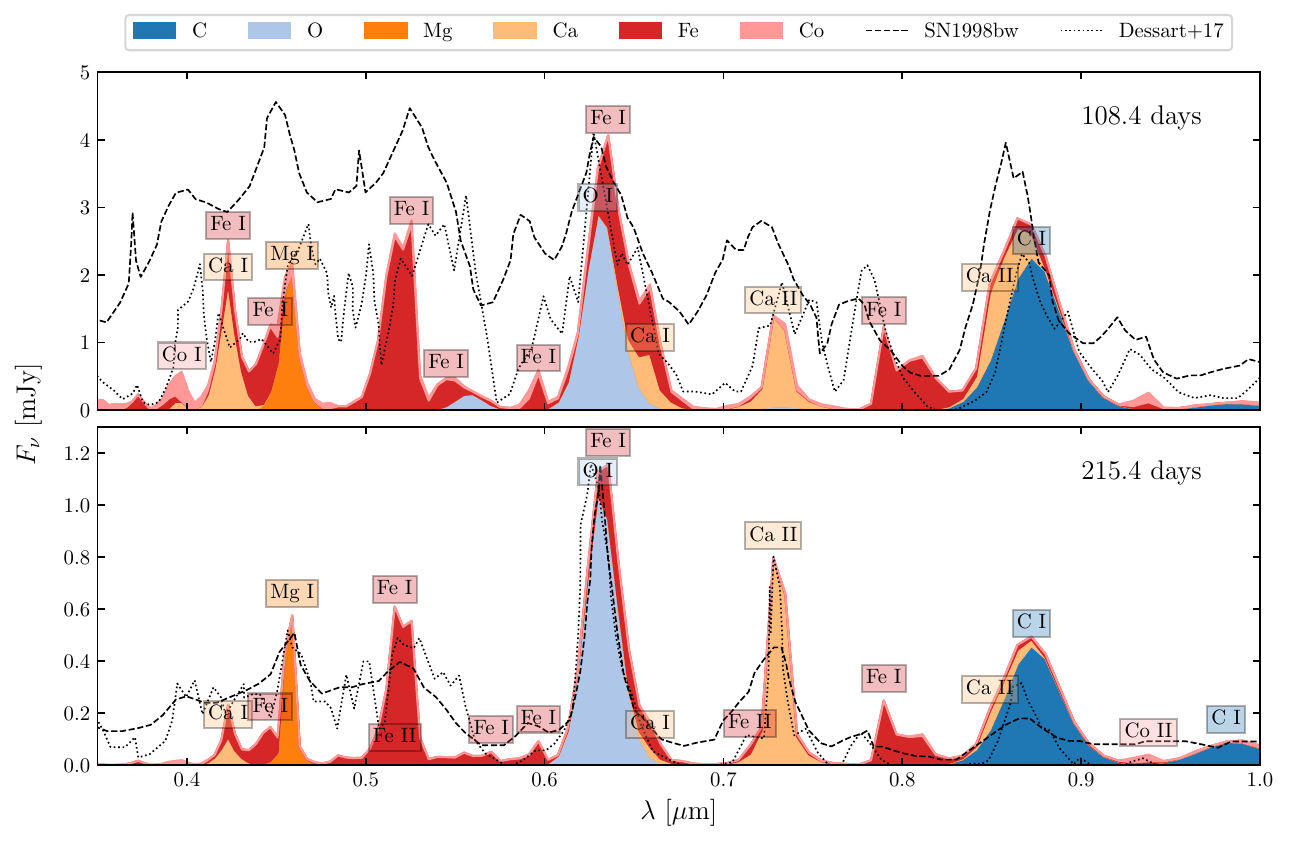}
    \caption{Optical to near-\ac{IR} spectra computed with our \ac{NLTE} model at 108.4 and 215.4 days post-explosion, compared to the observed spectra of \ac{SN} 1998bw, corrected for redshift and extinction (dashed line).
    The r0e2 model from \citet{Dessart:2017} is also shown for comparison.
    Both models are normalized to the observed feature at $\sim6300~\text{\AA}$ dominated by the O I forbidden doublet.
    The relevant contributions of the included elements to the total flux are shown by color-coding with respect to the element and stacking them according to increasing $Z$, while the specific ion species responsible for the features are explicitly noted.}
    \label{fig:spectra_epochs}
\end{figure*}

The nature of \ac{SN} 1998bw, at a luminosity distance of 37.8 Mpc \citep{Patat:2001}, was long debated, due to the intrinsic characteristics of its photometric and spectroscopic evolution.
Initially, \citep{Maeda:2003} proposed a stratified ejecta with a slow enough high density core and a fast enough outer shell, each containing \ce{^{56}Ni}, in order to explain the emission brightness both at early and late times.
This structure has been found difficult to be achieved in spherical symmetry, while it is consistent with more asymmetric explosions.
In fact, \citet{Dessart:2017} argued that such stratification is not even necessary, once the ejecta asymmetry is exploited.
The early phase of the emission could then be explained by a faster component with a relatively higher \ce{^{56}Ni} abundance in the polar region, while the later phase would result from a slower component with lower \ce{^{56}Ni} concentration along the equator.
In any case, the nebular phase is well described by a slow and massive ejecta with a limited amount of \ce{^{56}Ni}.
The latter is likely more indicative of the true element abundance, when compared to the incredibly highly energetic and \ce{^{56}Ni}-rich ejecta necessary to explain, in the spherically symmetric scenario, the first days of emission.
This conclusion from \citet{Dessart:2017} stems from the analysis of a series of synthetic \ac{SN} models, post-processed with the time-dependent 1D radiative transfer code CMFGEN \citep{Hillier:2012}.
The models were obtained by evolving a $40~M_\odot$ \ac{ZAMS} star with different metallicities and rotation degrees using the stellar evolution code MESA \citep{Paxton:2010,Paxton:2015}.
The resulting carbon-rich \ac{WR} stars were pushed to explosion with the radiation-hydrodynamics code V1D \citep{Livne:1993,Dessart:2009}, in order to obtain ejecta with different kinetic energies.

Here, we use as benchmark for our calculations the r0e2 model from \citet{Dessart:2017}, which derives from a \ac{ZAMS} star with no initial rotation and solar metallicity, eventually exploding with a kinetic energy of $4.12\times10^{51}$ erg, and resulting in the best fit of the late observations.
Our ejecta are constructed using three concentric zones, with the first representing the innermost region of the outflow, and the third the outermost.
We find that increasing further the number of zones does not lead to an improvement in the final prediction of the spectra, while, conversely, less zones also fail in capturing some of the emergent spectral features.
We focus on the observational epochs when we expect our model assumptions to be verified, around the first hundreds of days after the \ac{GRB}.
In particular, among the epochs analysed by \citet{Dessart:2017}, the one at $\sim215$ days appears to be the most suited to represent the conditions corresponding to pure emission nebula.

The main model parameters associated to each zone at this epoch are summarized in \reftab{tab:spectra_model}, and are directly informed on the r0e2 model.
We also show the correspondent composition in \reffig{fig:composition}.
We note that the mass values reported in \citet{Dessart:2017} only include the most relevant elements, such as those that produce observed spectral features.
Since the model r0e2 predicts a minor amount of He in the outer layers of the ejecta, we include this element when computing the ionization balance.
However, given its low mass, we do not include He lines in the emission model, since it is not observed to play a relevant role in the \ac{SN} 1998bw spectra.
We note that, in any case, generally the presence of a significant amount of He would be recognizable, assuming a Type Ib \ac{SN} event, through its most prominent emission features, such as the He I line at $5876~\text{\AA}$.
At longer wavelengths instead, He could be identified at earlier epochs by the He I P Cygni profiles at 1.08 and 2.06 ${\rm \mu m}$, as found in, e.g., the intermediate Type Ib/c \ac{SN} 1999ex \citep{Hamuy:2002kk}, or the \ac{SN} Ic 1994I \citep{Filippenko:1995}.
Conversely, since the amount of Mg is not reported in \citet{Dessart:2017}, we include manually this element in the composition, opting for an amount consistent with that of Si, and at the same time able to match the Mg I line at $4571~\text{\AA}$.
Since the resulting mass fractions do not add up to 1, we allocate the remaining ejecta mass to a representative average element, that we choose to be Ne.
The latter plays an almost inert role in the scope of our work.
In particular, it only produces a Ne II line at $12.81~{\rm \mu m}$, so that the correspondent power relative to the other elements is negligible.
Thus, the results of our analysis are not affected by its presence.

The slower innermost zone contains the majority of the total initial \ce{^{56}Ni} content, as in the original r0e2 model.
However, since our model considers non-thermal heating only from local sources, we slightly redistribute the initial \ce{^{56}Ni}, in order to have it in some minor amount also in the other zones, while keeping the total amount fixed to the r0e2 value of $\sim0.12~M_\odot$.
In this way, we roughly mimic the contribution to the emission coming from the outer zones, and powered by the inner one, without including non-local heating rates.
At the considered epoch, the energy deposition is governed by the \ce{^{56}Co} electron captures and $\beta^+$-decays, with the material being still opaque to $\gamma$-rays.
The latter, and in minor amount also primary positrons, release a few $10^7~{\rm erg~s^{-1}~g^{-1}}$ in the core of the ejecta.
As found by computing the ionization and thermal balance in each zone, this energy is spent to ionize matter more efficiently in the outer layer, due to the lower density.
The resulting kinetic temperature and ionization stage are the highest in the third zone, as shown in \reftab{tab:spectra_model} and \reffig{fig:composition}, respectively.
There, many species are relevantly ionized to the second stage, including Fe, and up to the third stage in the case of Mg and Ca.
Note that in this model the intermediate zone is instead the one with the lowest average ionization stage.
This is because the matter density is only slightly lower than the first zone, while the average atomic mass decreases relevantly, leading to a higher number density of atoms (see \reftab{tab:spectra_model}).
On the other hand, the lighter composition releases the input heating at higher temperatures, in absence of the strong Fe cooler, causing the kinetic temperature to monotonically increase from the core to the surface of the ejecta.

We compare in the lower panel of \reffig{fig:spectra_epochs} the resulting optical spectrum with the r0e2 model, and with the \ac{SN} 1998bw observation (corrected for extinction and redshift) at $\sim215$ days after the \ac{GRB} (\mbox{$\sim201$} days after B maximum light), from the coverage done by the \ac{ESO} on La Silla and \ac{VLT}/UT1 on Paranal \citep{Sollerman:2000,Patat:2001}.
The contributions to the emission coming from the different atomic species are displayed cumulatively, so that their combination is directly comparable with the r0e2 model and the observation.
Both the r0e2 and our model are rescaled to match the feature at $\sim6300~\text{\AA}$.
The latter, corresponding to the O I forbidden doublet at $6300-6363~\text{\AA}$, is the most prominent feature of this nebular spectrum, generated within the outer core of the ejecta (second zone in the model), and appearing already tens of days before the considered epoch.
The presence of such line, together with the narrower Mg I line at $4571~\text{\AA}$, and the Ca II doublet at $7300~\text{\AA}$ (produced in the first zone), are consistent with what observed in several spectra from Type Ib/c \acp{SN}, such as the \acp{SN} Ic 1987M and 1996aq, and the \ac{SN} Ib 1996N \citep[see, e.g.][for a direct comparison of these events]{Patat:2001}.
The relative strength of the above features is in agreement with what found by \citet{Dessart:2017} with the r0e2 model, even though both the synthetic spectra fail in reproducing the absolute strength of the Ca II line at $7300~\text{\AA}$.
Given the abundance of Ca II, a strong line from its triplet at $8498-8542-8662~\text{\AA}$ would be also expected, but it is missing, due to it being mostly populated through absorption.
Instead, we find a broad feature from the C I forbidden triplet at $8730~\text{\AA}$, that emerges from the outer ejecta layer (third zone).

Regardless of the Mg I line at $4571~\text{\AA}$, the region of bluer wavelengths $\lesssim5500~\text{\AA}$, which in our model is dominated by a series of Fe I lines from the ejecta core, is poorly reproduced.
Possibly, in this range at $\sim215$ days, there is still a residual continuum inherited from the past epochs, as it can be inferred by comparing the correspondent earlier spectra.
For example, the spectrum at $\sim108$ days (upper panel of \reffig{fig:spectra_epochs}) presents an evident underlying continuum, that extends up to the O I line and beyond.
As expected, our model is practically unable to reproduce the observed spectrum in such case, since it does not consider any continuum radiation, nor absorption effects other than self-absorption.
The absence of the latter effects, and in particular of Fe blanketing, could be a concurrent reason for the apparently overestimated Fe I emission flux, whereas, in the same wavelength range, \citet{Dessart:2017} argue for the presence of a forest of Fe II and Ti II lines, together with Ca II absorptions.
However, while singly ionized atoms are present in somewhat comparable amounts to the neutral atoms for most of the included elements, the absence of strong Fe II lines in particular also suggests a possible underestimation of the ionization processes.
In fact, we model ionization due to non-thermal electrons and recombination photons from the ground states, but we do not include ionization from the excited states, which could significantly enhance the ionization rates of neutral ions.
We investigate in \refsec{sec:discussion} the impact that our uncertainty on the degree of ionization has on our estimates for the r-process signatures in the spectra.

Despite the clear limitations due to the absence of an implementation of radiative transfer and the partial treatment of all the relevant atomic processes, the comparison with the spectra of \ac{SN} 1998bw highlights that our model is able to, at least approximately, reproduce some distinctive traits of late \ac{SN} Ib/c emissions.
We therefore take these features as references in order to explore the injection of species heavier than Fe group elements in the \ac{SN} ejecta.

\subsection{r-process signatures in the spectra}
\label{sec:r-process_signatures}

\begin{table}
    \centering
    \caption{Total energy flux on Earth from a \ac{SN} 1998bw-like event (at a luminosity distance of 37.8 Mpc, 215.4 days post-GRB), for different amounts of r-process elements mass $M_{\rm r}$ included in the composition.
    Fluxes are computed separately for the wavelength ranges $\leq1~{\rm \mu m}$ and $>1~{\rm \mu m}$, and their ratio $R_{F}=\frac{F_{\rm >1~\mu m}}{F_{\rm \leq1~\mu m}}$ is shown to quantify the effect of the inclusion of heavy elements.}
    \begin{tabular}{ccc|ccc}
    \toprule
    $M_{\rm r}$ & $\frac{M_{\rm r}}{M_{\rm ej}}$ & $\frac{M_{\rm r}}{M_{\ce{^{56}Ni}}}$ & $F_{\rm \leq1~\mu m}$ & $F_{\rm >1~\mu m}$ & $R_{F}$ \\
    & & & $\times10^{-13}$ & $\times10^{-13}$ & \\
    $[M_\odot]$ & & & $[{\rm erg~s^{-1}~cm^{-2}}]$ & $[{\rm erg~s^{-1}~cm^{-2}}]$ & \\
    \midrule
    0.00 & 0.000 & 0.00 & 7.22 & 0.54 & 0.08 \\
    0.03 & 0.003 & 0.25 & 6.40 & 0.73 & 0.11 \\
    0.06 & 0.007 & 0.50 & 5.65 & 0.90 & 0.16 \\
    0.12 & 0.013 & 1.00 & 4.88 & 1.23 & 0.25 \\
    \bottomrule
    \end{tabular}
    \label{tab:fluxes_ratio}
\end{table}

\begin{figure*}
    \centering
    \includegraphics[width=\textwidth]{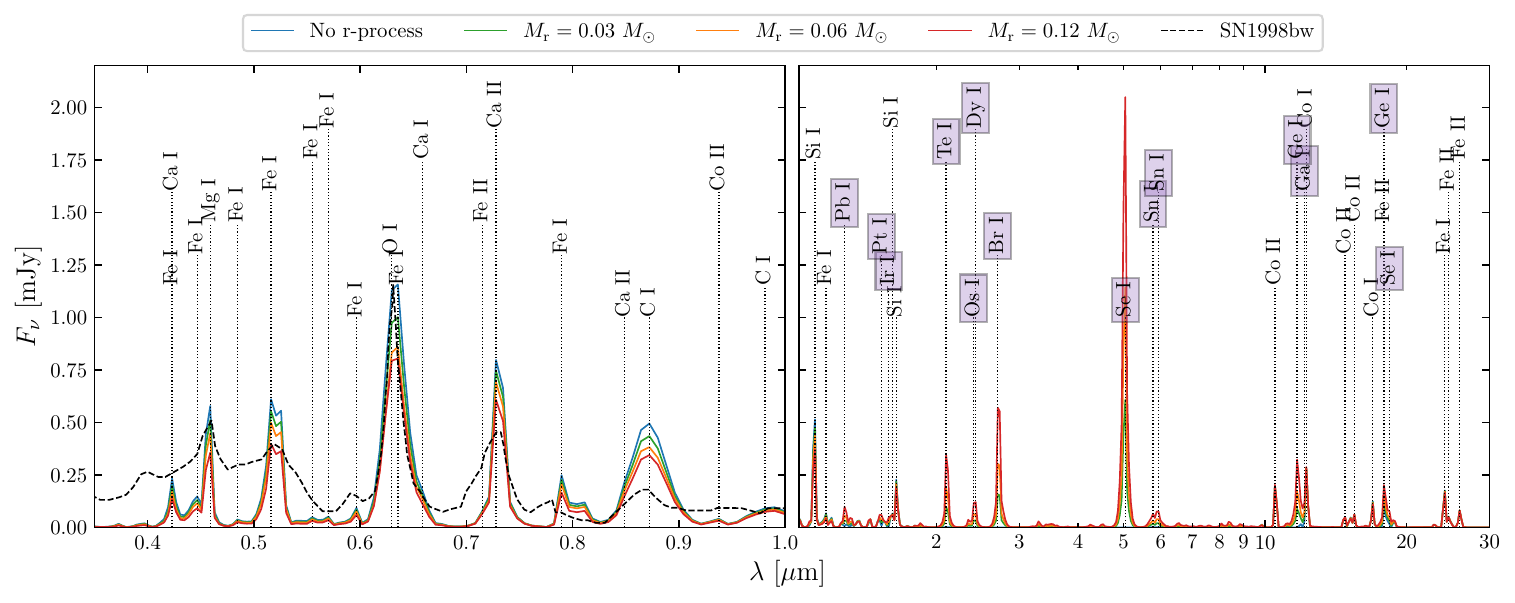}
    \caption{Optical to mid-\ac{IR} spectra computed with our \ac{NLTE} model targeted on \ac{SN} 1998bw at $215.4$ days post-explosion, for different amounts of r-process material included in the modeled composition.
    The observed \ac{SN} 1998bw spectra, corrected for redshift and extinction, is shown as reference (dashed line).
    The ion species responsible for the most relevant features are explicitly noted, with the new species with $Z\geq30$ highlighted by purple boxes.}
    \label{fig:spectra_r}
\end{figure*}

We start from the model constructed in \refsec{sec:reproducing_sn1998bw}, and we progressively introduce heavy elements ($Z>30$) into the composition.
For the relative abundance of such species, we adopt the solar r-process pattern from \citet{Prantzos2020}, and for simplicity we add the same mass of r-process elements to all the zones simultaneously, at the expenses of our inert element, i.e. Ne.
In this way, our modification of the composition does not affect the masses of the relevant lighter elements, nor their relative abundance.
We show in \reffig{fig:spectra_r} the spectra resulting from the inclusion of a mass of r-process elements $M_{\rm r}$ varying from zero to $0.12~M_\odot$, which is roughly the amount of the initial \ce{^{56}Ni} content.
In the absence of heavy elements, the emission flux is mostly concentrated at shorter wavelengths, in the optical region.
On the other hand, with our current transition list, heavy elements generate forbidden lines in the near/mid-\ac{IR}.
The near-\ac{IR} excess caused by their presence has been already suggested in several works before \citep[see, e.g.,][]{Siegel:2019}.
Here, in order to quantify it, we integrate the flux in the wavelength range $\leq1~{\rm \mu m}$, and the flux in the range $>1~{\rm \mu m}$, that is $F_{\rm \leq1~\mu m}$ and $F_{\rm >1~\mu m}$, respectively.
We compare the two fluxes by referring to their ratio $R_F=\frac{F_{\rm >1~\mu m}}{F_{\rm \leq1~\mu m}}$ in \reftab{tab:fluxes_ratio}.
Since we used Ne as our inert species, when computing the spectra (displayed in \reffig{fig:spectra_r}) and the total fluxes, we ignore the Ne II line at $12.81~{\rm \mu m}$.
In doing so, we assume that the residual mass which we do not account for does not change the values of $R_F$ significantly.

As visible in \reffig{fig:spectra_r}, in absence of elements heavier than iron group ones, we observe a flux ratio $R_F\sim10^{-1}$.
Specifically, besides the already mentioned features in the optical, in the near/mid-\ac{IR} we find lines of Fe I and Si I between 1 and 2 ${\rm \mu m}$, and lines of Fe I and II, and Co I and II, between 10 and 30 ${\rm \mu m}$.
The first new lines that appear, once a relatively small amount of heavy elements is included in the composition ($0.3\%$ of the total mass), are from Te I at $2.10~{\rm \mu m}$, Br I at $2.71~{\rm \mu m}$, Se I at $5.03~{\rm \mu m}$, and Ge I at $11.73~{\rm \mu m}$.
In particular, the line from Se I, here the strongest feature in the mid-\ac{IR} range, appears relevant also in recent \ac{KN} nebula calculations from \citet{Banerjee:2025eoh}.
The latter, as well as the ones from Ge I and Br I, are considerably difficult to interpret, due to the alternative nucleosynthetic pathways available for the production of such elements.
The consensus of the community on the subject is still in general disarray, whereas it is not clear if these elements can be created also in normal \acp{SN}.
Instead, the prompt appearance of the line from Te I (here with significant strength of $\sim0.1$ mJy at the considered distance), originating from the \ac{M1}+\ac{E2} transition $5p^4~\ce{^3P_1}~-~5p^4~\ce{^3P_2}$, would be a good indicator for an r-process which creates second peak elements.

A few times 0.01 $M_\odot$ of r-process material are sufficient for the Te I line at $2.10~{\rm \mu m}$ to have a strength comparable to that of the Si I lines at $1.10~{\rm \mu m}$ and $1.65~{\rm \mu m}$.
In Type Ic \ac{SN} events, such as \ac{SN} Ic 2007gr \citep{Hunter:2009tw}, or \ac{SN} Ic 2022jli \citep{Cartier:2024usw}, the above Si I features have been identified, while there is no evidence for similarly strong lines around $2.1~{\rm \mu m}$.
On this basis, we can thus place a constraint on the amount of r-process material produced in events of this kind.
We consider, for example, \ac{SN} 2007gr, which, at $\sim176$ days past B maximum light, shows a luminosity $\Delta_{\lambda}L_{\lambda}\sim9\times10^{37}~{\rm erg~s^{-1}}$ around $\sim2~{\rm \mu m}$ \citep{Hunter:2009tw}.
If we assume temperature and density conditions similar to those found in our model for \ac{SN} 1998bw ($T_{\rm e}\sim4000~{\rm K}$, $n_{\rm e}\sim10^7~{\rm cm^{-3}}$), we expect Te atoms to be mostly neutral and in the ground state, and the line at $2.1~{\rm \mu m}$ to be dominated by radiative de-excitations.
In this limit, the luminosity of the line can be approximated by
\begin{equation}
    \label{eq:Te_luminosity}
    \Delta_{\lambda}L_{\lambda} = N_{\rm Te} E_{\lambda} n_{\rm e} k_{\lambda} \, ,
\end{equation}
with $N_{\rm Te}$ the number of Te atoms, and $k_{\lambda}$ the collisional excitation rate coefficient of the line.
If we interpret the observed luminosity around $\sim2~{\rm \mu m}$ as a limit to the Te I line luminosity, we can invert \refeq{eq:Te_luminosity} to find a maximum amount of Te atoms, and by extension of r-process material:
\begin{equation}
    \label{eq:mass_limit}
    M_{\rm r} \lesssim \left(\frac{\Delta_{\lambda}L_{\lambda}}{5\times10^{41}~{\rm erg~s^{-1}}}\right) \left(\frac{X_{\rm r,Te}}{0.03}\right)^{-1} \left(\frac{n_{\rm e}\Omega_{\lambda}}{10^7~{\rm cm^{-3}}}\right)^{-1}~M_\odot \, .
\end{equation}
Here, $X_{\rm r,Te}$ is the mass fraction of Te relative to the r-process composition, with $0.03$ its value in the case of the solar r-process pattern \citep{Prantzos2020}, while $\Omega_{\lambda}$ is the collision strength of the Te I line ($\sim1$ in our calculations).
In the case of \ac{SN} 2007gr, assuming the values proposed in \refeq{eq:mass_limit}, we thus obtain a stringent limit to the r-process mass $M_{\rm r}\lesssim2\times10^{-4}~M_\odot$.

Injecting 0.06 $M_\odot$ of r-process content in our \ac{SN} 1998bw model composition causes the emergence of small features from Dy I and Pb I between 1 and 3 ${\rm \mu m}$.
Doubling this amount makes the flux in the same wavelength range to be enriched in small part also by Os I, Ir I, and Pt I lines.
These additional contributions from third peak elements appear only for a total mass of r-process material which is comparable to that of \ce{^{56}Ni}.
Since this is already a rather optimistic scenario, both in terms of r-process production and source distance, we expect that the potential detection of third peak r-process elements is in any case more difficult, compared to that of second peak elements.

Regardless, it appears clear that there is a systematic shift of the overall flux towards longer wavelengths, the more heavy elements are included (assuming our line list to be complete enough).
In the most extreme case considered, $F_{\rm >1~\mu m}$ reaches even a fourth of the optical part.
According to these estimates, we can therefore expect that the presence of heavy elements in the ejecta from this type of \acp{SN} may be already hinted by a value of $R_F$ which is greater than $\sim0.1$.
Clearly, the presence of r-process elements in events which are further away, compared to a few tens of Mpc, will prove more difficult to be deciphered.
However, the sensitivity limit of the new \ac{JWST} instrumentation in, e.g., $1-3~{\rm \mu m}$ with the NIRSpec and the NIRCam is of the order of $\sim0.1-10~{\rm \mu Jy}$, allowing for a much deeper search (up to hundreds of Mpc) of such revealing features in the spectra.

\subsection{Robustness of the results}
\label{sec:discussion}

\begin{figure}
    \centering
    \includegraphics[width=\columnwidth]{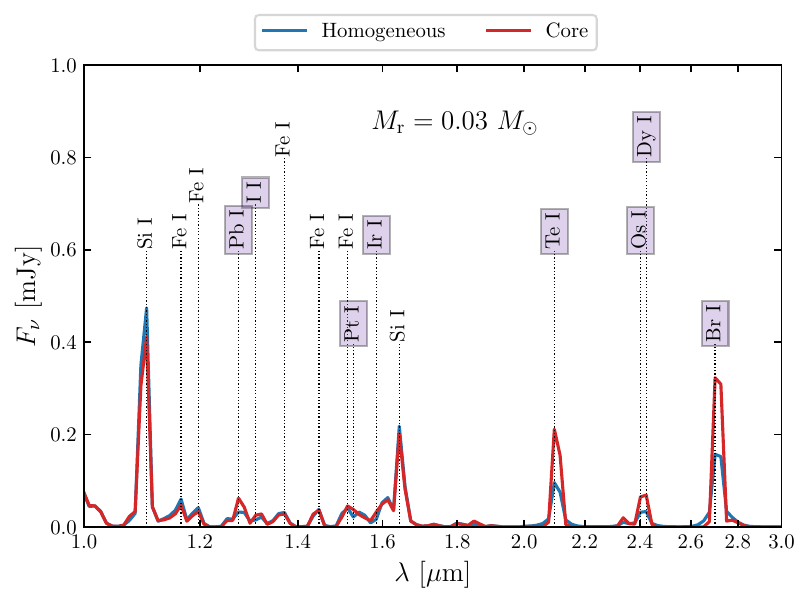}
    \caption{Zoom in the near-\ac{IR} of the spectrum shown in \reffig{fig:spectra_r} for the model with $M_{\rm r}=0.03~M_\odot$, obtained injecting heavy elements uniformly in all zones (homogeneous).
    For comparison, the model with the same mass of heavy elements concentrated in the inner zone (core) alone is also shown.}
    \label{fig:spectra_r_dist}
\end{figure}

\begin{figure}
    \centering
    \includegraphics[width=\columnwidth]{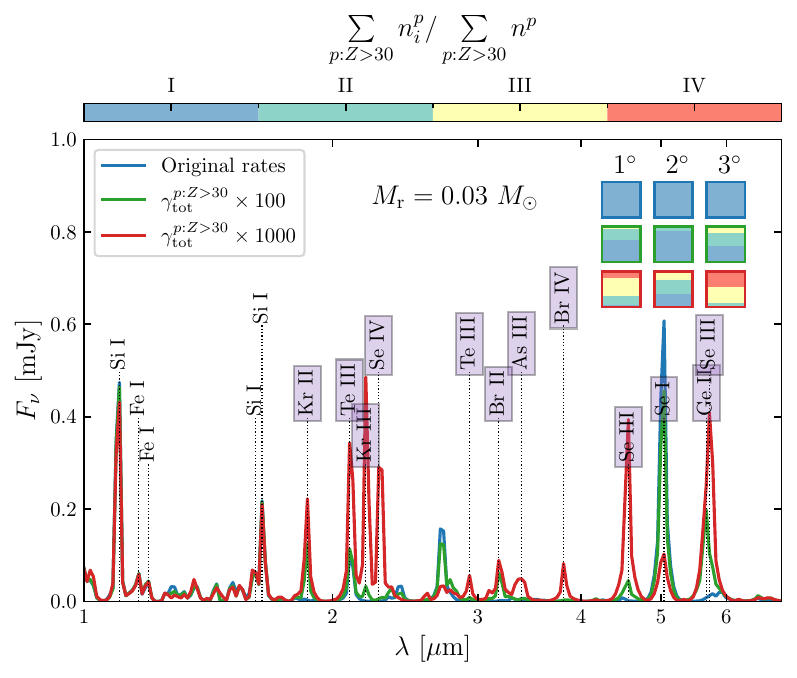}
    \caption{Zoom in the near to mid-\ac{IR} of the spectrum shown in \reffig{fig:spectra_r} for the model with $M_{\rm r}=0.03~M_\odot$, compared to the same model where the total ionization rate of heavy elements ($Z>30$) is systematically boosted of a factor 100, and 1000.
    For every model, the total ionization fractions of these heavy elements in each zone are shown as stacked histograms (top right of the figure).}
    \label{fig:spectra_r_ion}
\end{figure}

\begin{figure}
    \centering
    \includegraphics[width=\columnwidth]{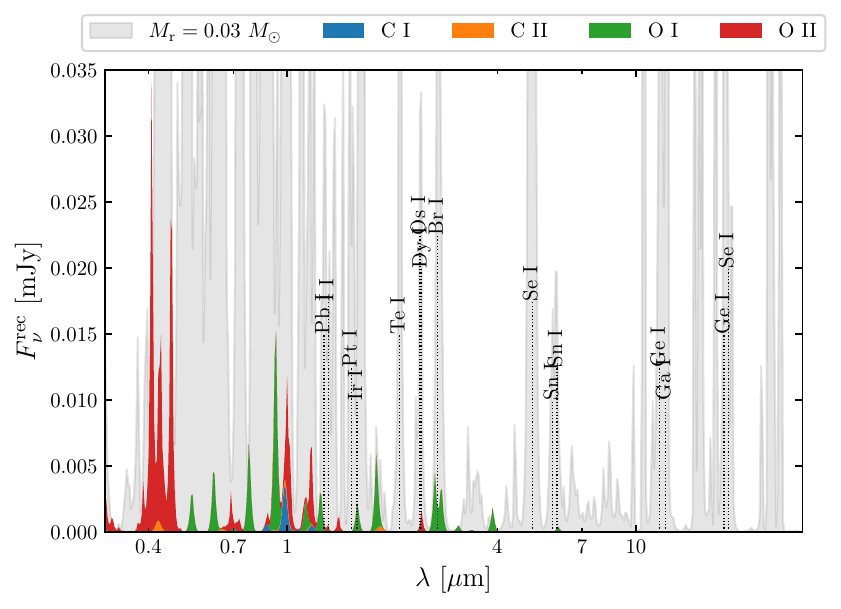}
    \caption{Recombination spectrum obtained for the composition and the ejecta conditions reported in \reffig{fig:composition} and \reftab{tab:spectra_model}, repectively.
    The contributions from the relevant ion species are color-coded and stacked to form the total flux.
    The grey background spectrum, and the corresponding most relevant heavy ion species noted, correspond to the model with $M_{\rm r}=0.03~M_\odot$, shown in \reffig{fig:spectra_r}.}
    \label{fig:spectra_rec}
\end{figure}

The argument exposed in \refsec{sec:r-process_signatures} on the detectability of heavy elements, in its generality, holds regardless of our assumptions on their distribution within the ejecta, as visible in \reffig{fig:spectra_r_dist} for our model with $M_{\rm r}=0.03~M_\odot$.
If we assume the entire amount of heavy elements to be contained within the inner zone alone, corresponding to the ejecta core, the only difference with a homogeneous distribution is a boost in the production of the same lines.
In fact, the ejecta conditions across the modeled zones are not sufficiently different to alter significantly the average ionization stage.
We note, however, that this conclusion is aided by the assumption that the medium is optically thin, whereas the emitted lines are not reprocessed in other ejecta regions.
Furthermore, we acknowledge that our treatment of ionization and recombination itself is affected by potentially large uncertainties, both due to the atomic data, and due to the atomic and thermodynamic processes neglected.
As seen in \refsec{sec:reproducing_sn1998bw}, the achieved ionization stage of lighter elements is in rough agreement with what it is observed in \ac{SN} 1998bw, despite some difficulties in accurately reproducing the high frequency features still remain.
On the other hand, the ionization state of heavier elements is less constrained, and radioactive elements may also be subject to some degree of clumping within the ejecta.

We thus explore how our findings would change, assuming the case in which our modeling, or alternative underlying ejecta conditions, would result in a different overall ionization stage of the system.
\reffig{fig:spectra_r_ion} shows the model with $M_{\rm r}=0.03~M_\odot$, where the total ionization rate for the heavy elements ($Z>30$) is increased artificially, first by a factor 100, and then 1000.
In the first case, singly ionized species appear in the composition in amounts comparable to the neutral species, with the most relevant new lines being from Kr II at $1.86~{\rm \mu m}$, Br II at $3.19~{\rm \mu m}$, and Ge II at $5.66~{\rm \mu m}$.
In the second case, boosting the ionization of heavy species further, we obtain a large part of the population to be in the third and even fourth ionization stage.
In such a scenario, we observe the Te I line at $2.10~{\rm \mu m}$ being substituted by the Te III line at the same wavelength, already found by \citet{Hotokezaka:2023aiq} in the nebular phase of the \ac{KN} AT2017gfo.
With similar strengths, also Kr III and Se IV produce observable lines, respectively at $2.20~{\rm \mu m}$ and $2.29~{\rm \mu m}$.
At longer wavelengths instead, Se III dominates with two lines at $4.55~{\rm \mu m}$ and $5.74~{\rm \mu m}$, taking the place of the Se I line at $5.03~{\rm \mu m}$.
Furthermore, additional smaller features are generated by Te III, As III and Br IV, at $2.93~{\rm \mu m}$, $3.40~{\rm \mu m}$, and $3.81~{\rm \mu m}$, respectively.

Compared to the results obtained with the original ionization rates, we note that the different ionization stage of the system results in the most prominent features from neutrals being substituted by similarly strong features from their ionized counterparts, at similar wavelengths.
In particular, the elements originally producing the most relevant lines continue to characterize the near to mid-\ac{IR} spectrum, with the possible addition of Kr and As.
As such, the spectrum remains shaped by the lighter heavy elements with $Z=32-36$, which are co-produced via different channels, and by Te, which continues to be potentially detectable, assuming the material in the line forming region at the considered epoch being not too fast.

On a different note, however, we have to consider the possibility that the wavelength region of interest could be in general polluted by additional lines emerging from recombination photons.
For this reason, here we give an estimate for their potential presence, using data available in literature.
We model the recombination spectrum in a similar way to \refeq{eq:spectrum}, that is
\begin{equation}
    \begingroup
    \mathcode`v=\varv
    L_\nu = \sum_{m,l:m>l} N_m E_{ml} n_{\rm e} \alpha_{ml} \mathcal{N}_\nu\left(\nu_{ml},\left(\frac{\nu_{ml}v}{\sqrt{2}c}\right)^2\right) \, ,
    \endgroup
\end{equation}
where the sum runs over the lines emitted from excited levels $m$ into which the atoms recombine with a rate $\alpha_m$.
The latter is used to compute the line rate coefficient
\begin{equation}
    \alpha_{ml} = \frac{A_{ml}}{\sum\limits_{n:m>n}A_{mn}} \alpha_{m} \, ,
\end{equation}
which accounts for the branching to different lower states.
We include the relevant lines from the species responsible for most of the recombination events, by importing level-specific recombination rates for C I from \citet{Nahar:1995}, C II from \citet{Davey:2000}, O I from \citet{Nahar:1999}, O II from \citet{Storey:1994}, and Mg I and Si I from \citet{Nussbaumer:1986}.

As shown in \reffig{fig:spectra_rec} for the model with $M_{\rm r}=0.03~M_\odot$, C and O are the only relevant contributers to the recombination spectrum, with O I being the dominant source in the optical region, while the near-\ac{IR} being mainly occupied by O I and II lines.
Using the parameters of the zones in the models considered above, we obtain a total recombination photon flux which is more than one order of magnitude smaller than the flux of photons from collisionally excited atoms, both in the optical and in the \ac{IR} range.
In this regard, we note that shifts in the current lines, as well as the appearance of additional lines, could potentially emerge from the employment of a more complete and precise dataset.
However, we expect the overall strength of such features to be comparable with our current estimates, given the same ejecta conditions.
This result justifies the fact that we do not calculate the spectrum of recombination photons including it in the overall thermal balance of the system, and suggests that the lines from heavy elements in the near/mid-\ac{IR} range should remain detectable.

\section{Conclusion}
\label{sec:conclusion}

Here we proposed an estimate of the impact that r-process elements are expected to have in the emergence of the optical to mid-\ac{IR} nebular emission from the ejecta of a \ac{SN} event.
The latter must be peculiar enough in order to harbor the nucleosynthesis of such species, as it could potentially occur in a \ac{MRSN} or a collapsar.
Among the different classes of \acp{SN}, \ac{GRB} \ac{SN} (\ac{BL} Type Ic) point to the above scenario, being more energetic, and featuring a relativistic jet.
We therefore referred to a representative event, \ac{SN} 1998bw, modeling its nebula a few hundreds days post-explosion.
The model employed solves the ionization, atomic level, and thermal balance, in an approximate way, by assuming the ejecta to be optically thin throughout its depth, and in a steady-state regime, where atomic processes are fast compared to the dynamical timescale.
We therefore study the epochs that proceed from the photospheric phase of the explosion, but are well contained within the first couple of years.
This simplification allows for the inclusion of an arbitrary number of elemental species, which we characterize using the atomic data mostly available in public databases.
Unfortunately, for a list of elements including r-process products, such data are still largely incomplete, which introduces a further source of uncertainty in the model.
However, the generality of our estimates makes them rather independent on the quality of the atomic data employed, whereas variations, e.g., on the recombination rates should lead to similar detectability conclusions.

Starting from these assumptions, we first explored the overall role of r-process elements in heating and cooling the ejecta.
We noted that for typical masses of \ce{^{56}Ni} and \ce{^{44}Ti}, consistent with the kind of explosions considered, $\sim 10^{-2}~M_\odot$ of r-process material could potentially leave a mark only at a very late stage of the light curve, around $\sim10$ years, in the form of a different decay slope.
However, this evidence could be very well polluted by the presence of additional energy sources, and to a minor extent also by the less constrained \ce{^{57}Ni}.
On the other hand, at least in the steady-state phase of the nebula, we found that the same amount of heavy elements is not likely sufficient to impact its cooling process.
Despite the greater cooling efficiency expected for single heavy elements such as Nd, the large presence of lighter elements like O, Ca, and up to Fe, requires the heavier species to constitute at least $\sim1\%$ of the total ejected mass, in order to contribute relevantly to the cooling rate.
In addition, a non trivial dependence on the kinetic temperature is retained, due to the different atomic structure and consequent transition lines of such species.

However, targeting the representative case of \ac{SN} 1998bw, we also found that the near to mid-\ac{IR} range of the emission spectrum leaves room for heavy elements to be possibly detected regardless, due to the both minor and cleaner emission from lighter species, compared to the optical range.
This result appears rather solid even in relation to the potential pollution by lines from recombined ions, and for any heavy elements distribution within the ejecta.
For example, in the specific case considered, a few times $10^{-2}~M_\odot$ of solar-type r-process material (corresponding to a few $0.1\%$ of the ejecta mass) could leave a signature in the form of a line of Te I at $2.10~{\rm \mu m}$.
Thus, this feature could be used as smoking gun for the occurrence of a relatively strong r-process, or, conversely, its absence could be used to constrain the produced amount of r-process elements.
For example, in the case of \ac{SN} 2007gr, we estimated an r-process mass limit of $M_{\rm r}\lesssim2\times10^{-4}~M_\odot$.
Alternatively, a line from Te III at the same wavelength could still serve the same purpose, in the hypothesis that the uncertainties in our modeling would substantially affect our estimates, and we would strongly underestimate the overall ionization stage. 
On a different note, independently from the identification of single elemental species, a general hint for the presence of r-process elements could be the ratio between near to mid-\ac{IR} and optical flux, because of the overall asymmetry in the cooling frequencies of light and heavy elements.
We estimated such ratio to be above $\sim0.1$, in order to be significantly linkable to the presence of heavy elements.

In conclusion, considering the new observational capabilities of \ac{JWST}, we expect the detection of r-process elements to be realistically feasible, especially by focusing the search efforts in a wavelength interval comprised between $\sim1-10~{\rm \mu m}$.
The presence (or absence) of such species in the ejecta from special type of \acp{SN} would then provide a tremendous assist to the comprehension of these events, shedding light over the mechanisms responsible for the explosion and the subsequent nucleosynthesis.
Ultimately, \ac{JWST} observations will prove essential for the understanding of the origin of heavy elements in the universe.

\section*{Acknowledgements}

We thank Anders Jerkstrand, Brian Metzger, and Masaomi Tanaka for useful discussions.
This work is supported by the Deutsche Forschungsgemeinschaft (DFG, German Research Foundation, Project ID 279384907, SFB 1245), the State of Hesse within the Research Cluster ELEMENTS (Project ID 500/10.006), the JST FOREST Program (JPMJFR2136), and the JSPS Grant-in-Aid for Scientific Research (20H05639, 20H00158, 23H01169, 23H04900).
CHIANTI is a collaborative project involving George Mason University, the University of Michigan (USA), University of Cambridge (UK), and NASA Goddard Space Flight Center (USA).

\section*{Data availability}

The data concerning this work will be shared on reasonable request to the authors.

\bibliographystyle{mnras}
\bibliography{refs,local}

\appendix

\section{Data for energy levels, radiative transition rates and collision strengths}
\label{app:data}

Energy levels are taken from the \ac{NIST} ASD \citep{NIST_ASD}.
\begin{itemize}
\item For iron group elements ($Z=26,27,28$), the radiative transition rates are taken from the \ac{NIST} database for Fe I, IV, Co I, Ni I, II, III, and IV, while from \cite{Nahar1995A&A} for Fe II, \cite{Nahar1996A&AS} for Fe III, \cite{Storey2016MNRAS} for Co II, and \cite{Storey2016MNRAS2} for Co III.
For Co I and IV, we calculate \ac{M1} transition rates using the analytic formula based on LS coupling \citep{Pasternack1940ApJ,Shortly1940,Bahcall1968ApJ}:
\begin{equation}
    \label{eq:M1}
    A_{ul}= 1.3\,{\rm s^{-1}}\,\left(\frac{\lambda_{ul}}{4\,{\rm \mu m}} \right)^{-3}f(J_u,L_u,S_u) \, ,
\end{equation}
where $\lambda_{ul}$ is the line wavelength for the transition between the upper level $u$ and lower level $l$, and $f(J,L,S)$ is an algebraic factor, i.e.
\begin{equation}
    f(J,L,S) = \frac{(J^2-(L-S)^2)((L+S+1)^2-J^2)}{12J(2J+1)}
\end{equation}
for $J_u=J_l+1$, and
\begin{align}
    f(J,L,S) = \frac{((J+1)^2-(L-S)^2)((L+S+1)^2-(J+1)^2)}{12(J+1)(2J+1)}
\end{align}
for $J_u=J_l-1$, where $J$ is the total angular momentum, $L$ the total orbital angular momentum, and $S$ the total spin angular momentum, in units of $\hbar$.
Collision strength data are taken from \cite{Pelan1997A&AS} for Fe I, \cite{Zhang1995A&A} for Fe II, \cite{Zhang1996A&AS} for Fe III, \cite{Zhang1997A&AS} for Fe IV, \cite{Storey2016MNRAS} for Co II, \cite{Storey2016MNRAS2} for Co III, and \cite{Bautista2004A&A} for Ni II.
The collision strengths for the lowest levels of the ground configuration of Ni III are taken from \cite{Blondin2023A&A}.
We use the collision strengths of Co III for Ni IV.
For the missing iron group collision data, we use the approximate formula from \citet{vanregemorter:1962} for \ac{E1} transitions, and a constant value $\Omega_{\rm F}$ for forbidden transitions.
We choose $\Omega_{\rm F} = 0.5$ for such elements.

\item For light elements ($Z\leq 25$), we take from the \ac{NIST} database the radiative transition rates for C I, II, N I, II, III, O I, II, III, IV, F I, II, III, IV, Ne II, III, Na I, III, IV, Mg I, II, IV, Al I, II, Si I, II, P I, II, III, S I, II, III, IV, Cl I, II, III, IV, Ar II, III, IV, K I, III, IV, Ca I, II, IV, Sc I, II, Ti I, II, V I, II, IV, Cr II, IV, and Mn I.
We use equation \ref{eq:M1} to obtain \ac{M1} transitions for C III, IV, N IV, Ne IV, Al III, IV, Si III, IV, P IV, Sc III, IV, Ti III, IV, V III, Cr I, III, Mn II, III, and IV.
We use the collision strengths given by \citep{Bhatia1995ApJS} for O I, \citep{Mao2020A&A} for O II, \citep{Storey2016MNRAS} for O III, \citep{Merle2015A&A} for Mg I, \citep{Sigut1995JPhB} for Mg II, \citep{Tayal1996JPhB} for Ar II, \citep{Ariii1998A&AS} for Ar III, \citep{CaII2007A&A} for Ca II, and \citep{Pelan1995A&AS} for Ca IV.
The remaining collision data for light elements are again computed with the \citet{vanregemorter:1962} formula, and $\Omega_{\rm F} = 0.5$.

\item For heavy elements ($Z\geq29$), we integrate the radiative \ac{M1} transitions from the \ac{NIST} database with the transitions from \citet{Hotokezaka:2022}, computed using equation \ref{eq:M1}, while for \ac{E1} transitions we exploit the \texttt{HULLAC} calculations carried by \citet{Hotokezaka:2021}.
In particular, we take the Nd \ac{E1} cooling power from that work, and we use it to characterize all the heavy elements.
In order to take into account the variable intensity of such transitions across the periodic table, we associate $10\%$ of the original \ac{E1} intensity to each element.
However, we do not include such \ac{E1} transitions individually when deriving the spectrum.
For the collision strength of transitions between levels of the ground terms, we use the data computed by \citet{Hotokezaka:2021} with \texttt{HULLAC}.
To treat the remaining collision strengths, we use the \citet{vanregemorter:1962} formula for \ac{E1} transitions, by assuming that radiative transition rates are approximately $A\approx2\times10^6\lambda^{-3}~{\rm s^{-1}}$, with $\lambda$ the corresponding wavelength in ${\rm \mu m}$.
Following \citet{Hotokezaka:2021}, we choose $\Omega_{\rm F}=1$ for forbidden transitions.
\end{itemize}

We note that recently new, more accurate atomic data targeted on elements relevant for \acp{KN} have been published, including collisional strengths for Sr and Y \citep{Mulholland:2024bwb}, and for Te \citep{Mulholland:2024cft}.
We plan on employing such data, together with further developments on our code, in future works, in order to refine our predictions on the detectability of heavy elements.

\section{Impact of photoionization from recombination photons}
\label{app:photoionization_impact}

\begin{figure*}
    \centering
    \includegraphics[width=\textwidth]{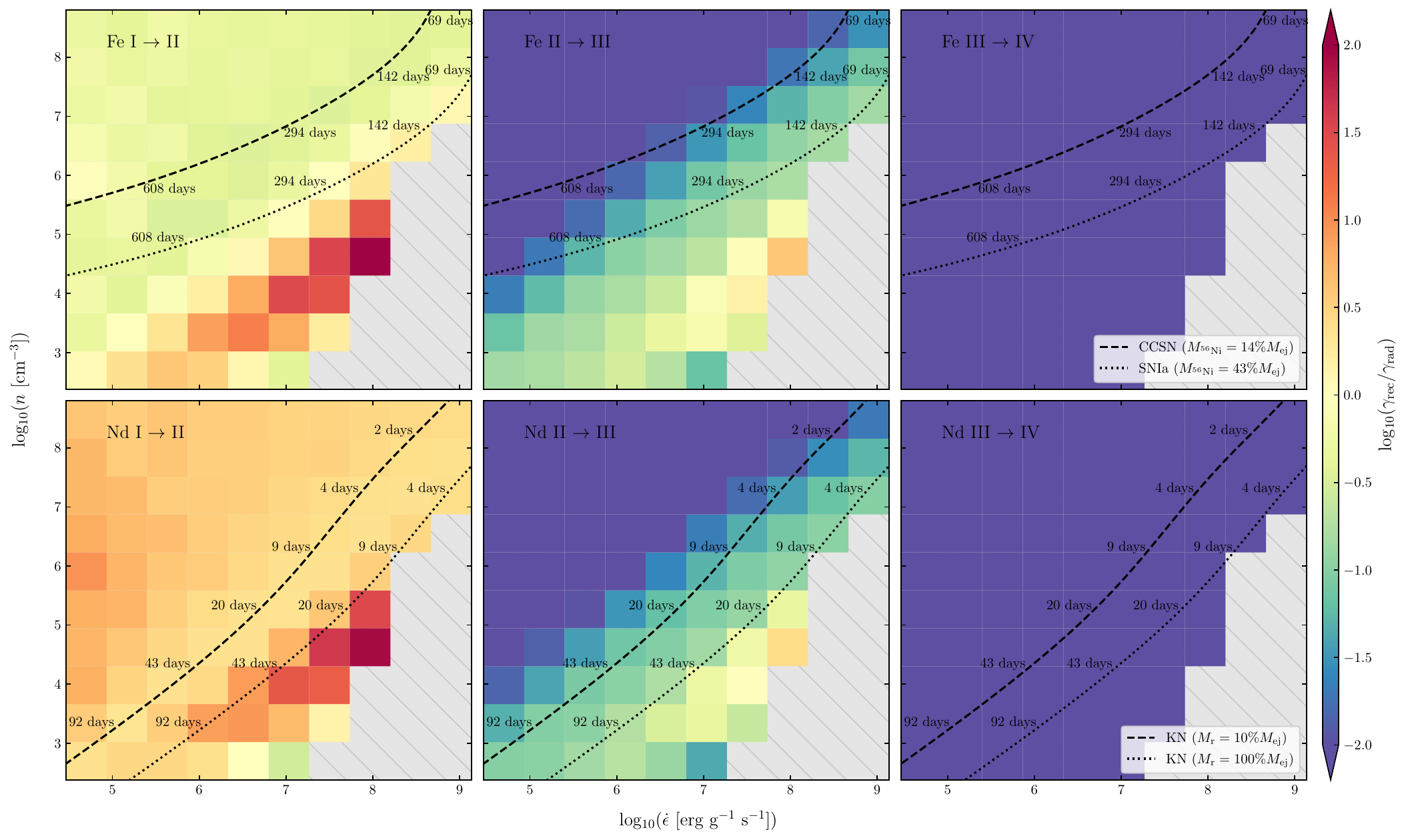}
    \caption{Ratio between the contribution to the ionization rate per atom coming from recombination photons and the contribution coming from decay $\beta$-particles, as a function of the specific energy deposition rate and the number density of atoms.
    The rates are computed for a mixture of $50\%$ Fe I-IV and $50\%$ Nd I-IV in mass fraction, on a 10x10 grid log-spaced, using ionization and thermal balance.
    Each panel shows a different ionizing species.
    For approximate visual reference, we superimpose to Fe rates (upper panels) plausible trajectories with their time evolution of a \ac{CCSN} and a \ac{SN} Ia, while we consider a \ac{KN} with two degrees of r-process material dilution for Nd rates (lower panels).}
    \label{fig:grid_photoion}
\end{figure*}

We test the relevance of the contribution to ionization coming from recombination photons $\gamma_{\rm rec}$ by considering the model parameter space explored in \refsec{sec:model_applicability}.
\reffig{fig:grid_photoion} shows the ratio between such contribution and that provided by non-thermal $\beta$-particles coming from radioactive decays $\gamma_{\rm rad}$, for all the atomic species in the considered mixture.
The results are displayed for the equal-mass Fe-Nd case, but they qualitatively hold also for the pure Fe and pure Nd cases, and by extension for other generic compositions.
The importance of this ionization mechanism heavily depends on the ionization stage of the targeted atomic species, while it is less dependent on the element itself.
For a high stage of ionization, such as III $\rightarrow$ IV, photoionization is absent, since only higher energy photons with a few tens eV can ionize the target ions.
On the contrary, most of the recombination photons lie in an energy range one order of magnitude lower.
As a result, the higher impact is observed for the ionization of stage I $\rightarrow$ II, while the stage II $\rightarrow$ III is only partially impacted, mostly by photons from the recombination of stage IV $\rightarrow$ III.
This effect has thus a clear dependence on the kinetic temperature, whereas such higher stages are most likely to be found at higher $T_{\rm e}$ (see also \reffig{fig:grid_Te} and \reffig{fig:grid_frac}).
As a result, for most of the explored parameter space, the ionization of neutral Fe is driven in comparable measure by the two modeled sources of ionization.
At lower densities and higher values of input heating, the contribution from recombination photons can even surpass the one from radioactivity by one order of magnitude or more.
This effect is even more relevant for the neutral Nd case, in which the photoionization dominates for the entirety of the parameter space.
In this regard, however, we note that these estimates are obtained by employing a certain number of assumptions on the photoionization cross-sections of the different ions, and by constructing a very rough recombination photon spectrum.
Therefore, despite the generality of the qualitative trends, we expect potential quantitative changes in the magnitude of the photoionization impact, once better treatments for this process are included.

\bsp    
\label{lastpage}
\end{document}